\documentclass[usenatbib]{mnras}
\usepackage{graphicx,amsmath,amssymb}
\usepackage{natbib}
\usepackage[english]{babel}

\def \etal {et~al.~}

\newcommand{\hMpc}{{\ifmmode{h^{-1}{\rm Mpc}}\else{$h^{-1}$Mpc}\fi}}
\newcommand{\Mpc}{{\ifmmode{{\rm Mpc}}\else{Mpc}\fi}}
\newcommand{\hkpc}{{\ifmmode{h^{-1}{\rm kpc}}\else{$h^{-1}$kpc}\fi}}
\newcommand{\kpc}{{\ifmmode{ {\rm kpc} }\else{{\rm kpc}}\fi}}
\newcommand{\kms}{{\ifmmode{ {\rm km\,s^{-1}} }\else{ ${\rm km\,s^{-1}}$ }\fi}}
\newcommand{\hMsun}{{\ifmmode{h^{-1}{\rm {M_{\odot}}}}\else{$h^{-1}{\rm{M_{\odot}}}$}\fi}}
\newcommand{\Msun}{{\ifmmode{{\rm M}_{\odot}}\else{${\rm M}_{\odot}$}\fi}}
\newcommand{\Mhalo}{{\ifmmode{M_{\rm halo}}\else{$M_{\rm halo}$}\fi}}
\newcommand{\Rvir}{{\ifmmode{R_{\rm vir}}\else{$R_{\rm vir}$}\fi}}
\newcommand{\Mvir}{{\ifmmode{M_{\rm vir}}\else{$M_{\rm vir}$}\fi}}
\newcommand{\Mstar}{{\ifmmode{M_{\rm star}}\else{$M_{\rm star}$}\fi}}
\newcommand{\Vrot}{{\ifmmode{V_{\rm rot}}\else{$V_{\rm rot}$}\fi}}
\newcommand{\ltsima}{$\; \buildrel < \over \sim \;$}
\newcommand{\gtsima}{$\; \buildrel > \over \sim \;$}
\newcommand{\lsim}{\lower.5ex\hbox{\ltsima}}
\newcommand{\gsim}{\lower.5ex\hbox{\gtsima}}

\def\lesssim{\mathrel{\hbox{\rlap{\hbox{\lower4pt\hbox{$\sim$}}}\hbox{$<$}}}}
\def\gtrsim{\mathrel{\hbox{\rlap{\hbox{\lower4pt\hbox{$\sim$}}}\hbox{$>$}}}}

\newcommand{\beq}{\begin{equation}}
\newcommand{\eeq}{\end{equation}}
\def\beqa{\begin{eqnarray}}
\def\eeqa{\end{eqnarray}}
\def\LCDM{\ensuremath{\Lambda}CDM}

\def\head{ \vbox to 0pt{\vss \hbox to 0pt{\hskip 440pt\rm
      LA-UR-10-07069\hss} \vskip 25pt}}

\def \kms {\ifmmode  \,\rm km\,s^{-1} \else $\,\rm km\,s^{-1}  $ \fi }
\def \kpc {\ifmmode  {\rm kpc}  \else ${\rm  kpc}$ \fi  }  
\def \hkpc {\ifmmode  {h^{-1}\rm kpc}  \else ${h^{-1}\rm kpc}$ \fi  }  
\def \hMpc {\ifmmode  {h^{-1}\rm Mpc}  \else ${h^{-1}\rm Mpc}$ \fi  }  
\def \Mpch {\ifmmode  {h^{-1}\rm Mpc}  \else ${h^{-1}\rm Mpc}$ \fi  }  
\def \Msun {\ifmmode {\rm M}_{\odot} \else ${\rm M}_{\odot}$ \fi} 
\def \hMsun {\ifmmode h^{-1}\,\rm M_{\odot} \else $h^{-1}\,\rm M_{\odot}$ \fi}

\def \LCDM {\ifmmode \Lambda{\rm CDM} \else $\Lambda{\rm CDM}$ \fi}
\def \sig8 {\ifmmode \sigma_8 \else $\sigma_8$ \fi} 
\def \OmegaM {\ifmmode \Omega_{\rm m} \else $\Omega_{\rm m}$ \fi} 
\def \Omegab {\ifmmode \Omega_{\rm b} \else $\Omega_{\rm b}$ \fi} 
\def \OmegaL {\ifmmode \Omega_{\rm \Lambda} \else $\Omega_{\rm \Lambda}$\fi} 
\def \Deltavir {\ifmmode \Delta_{\rm vir} \else $\Delta_{\rm vir}$ \fi}
\def \rhocrit {\ifmmode \rho_{\rm crit} \else $\rho_{\rm crit}$ \fi}
\def \rhou {\ifmmode \rho_{\rm u} \else $\rho_{\rm u}$ \fi}
\def \zc {\ifmmode z_{\rm c} \else $z_{\rm c}$ \fi}

\def\apj{ApJ}%
\def\apjl{ApJ}%
\def\apjs{ApJS}%
\def\ao{Appl.~Opt.}%
\def\aap{A\&A}%
\def\mnras{MNRAS}%
\def\na{New A}%
\def\pasp{PASP}%
\def\nat{Nature}%
\def\head{ .ps \vbox to 0pt{\vss \hbox to 0pt{\hskip 440pt\rm
      LA-UR-10-07069\hss} \vskip 25pt}} 

\def \spose#1{\hbox  to 0pt{#1\hss}}  
\def \lta{\mathrel{\spose{\lower 3pt\hbox{$\sim$}}\raise 2.0pt\hbox{$<$}}}
\def \gta{\mathrel{\spose{\lower 3pt\hbox{$\sim$}}\raise 2.0pt\hbox{$>$}}}

\title[NIHAO XIII - Clumpy galaxies]
{NIHAO XIII: Clumpy discs or clumpy light in high redshift galaxies?}

\author[T. Buck \etal] {Tobias Buck$^{1,2}$\thanks{E-mail: buck@mpia.de},
    Andrea V. Macci\`o$^{3,1}$\thanks{E-mail: maccio@nyu.edu},
    Aura Obreja$^{3}$,
    Aaron A. Dutton$^{3}$,
    \newauthor{Rosa Dom\'inguez-Tenreiro$^{4}$,
    Gian Luigi Granato$^{5}$}
    \\
$^1$Max-Planck-Institut f\"ur Astronomie, K\"onigstuhl 17, 69117 Heidelberg, Germany\\
$^2$Member of the International Max Planck Research School for Astronomy and Cosmic Physics at the University of Heidelberg,\\ 
\,\,IMPRS-HD, Germany\\
$^3$New York University Abu Dhabi, PO Box 129188, Saadiyat Island, Abu Dhabi, United Arab Emirates\\
$^4$Depto. de F\'isica Te\'orica, Universidad Aut\'onoma de Madrid, E-28049 Cantoblanco Madrid – Spain\\
$^5$INAF, Osservatorio Astronomico di Trieste, Via Tiepolo 11, I-34131 Trieste – Italy\\
}

\begin{document}

\date{Accepted 2017 March 16. Received 2017 March 16; in original form 2016 December 15}

\pagerange{\pageref{firstpage}--\pageref{lastpage}} \pubyear{2016}

\maketitle

\label{firstpage}


\begin{abstract}
Many massive star forming disc galaxies in the redshift range 3 to 0.5 are observed to 
have a clumpy morphology showing giant clumps of size $\sim$1 kpc 
and masses of about $10^7\Msun$ to $10^{10} \Msun$. The nature 
and fate of these giant clumps is still under debate.
In this work we use  19 high-resolution simulations of disc galaxies
 from  the NIHAO  sample  to  study the  formation  and the evolution  of clumps  in the  discs  of  high redshift  galaxies.
We use mock HST - CANDELS observations created with the radiative transfer code GRASIL-3D to carry out, for the first time, 
a quantitative comparison of the observed fraction of clumpy galaxies and its evolution with redshift with simulations. We find a good agreement between the observed clumpy fraction and the one of the NIHAO galaxies.
We find that dust  attenuation can suppress intrinsically  bright clumps and enhance less luminous ones. 
In our galaxy sample we only find clumps in light (u-band) from young stars but not in stellar mass surface density maps. This means
    that the NIHAO sample does not show clumpy stellar discs but rather a clumpy light distribution originating from clumpy star formation events. The clumps found in the NIHAO sample match observed age/color gradients as a function of distance from the galaxy center but they show no sign of inward migration. Clumps in our simulations disperse on timescales of a about a hundred Myr and their contribution to bulge growth is negligible.
\end{abstract}

\noindent
\begin{keywords}

  galaxies:  -  formation  - galaxies: - high-redshift -
  galaxies: evolution - galaxies: bulges - galaxies: ISM - 
  methods: numerical

 \end{keywords}

\vspace{1cm}
\section{Introduction}
\label{sec:introduction}

\begin{table*}
\label{tab:sims}
\begin{center}
\caption{Properties of the high mass end ($M_{\rm star}>10^9$ M$_{\odot}$) of
  NIHAO  galaxies  at  $z~1.5$.  The first column contains the name of the galaxy, column 2 to 4 state the number of particles within $R_{200}$
  all species, dark matter and stars. Column 5 shows the total mass within $R_{200}$ of the galaxy, 
  column 6 gives the stellar mass measure within 0.2$R_{\rm 200}$, column 7 is the value of $R_{200}$, 
  column 8 states the gas mass within 0.2$R_{\rm 200}$ and column 9 shows the fraction of cold gas ($<30000$ K) 
  and the last column shows the SFR measured within 0.2$R_{\rm 200}$.}
\begin{tabular}{l l l l l l l l l l}
\hline
\hline
Name & $N_{200}$ & $N_{\rm dark}$ & $N_{\rm star}$ & $M_{200}$ & $M_{\rm star}$ & $R_{200}$ & $M_{\rm gas}$ & $f_{\rm cold gas}$ & SFR \\
\hline
g2.04e11 & 1,122,554 & 635,540 & 110,861 & $1.53\times10^{11}$ & $9.01\times10^{8}$ & $62.74$ & $5.03\times10^{9}$ & $0.77$ & 0.61 \\ 
g2.41e11 & 814,971 & 467,054 & 55,838 & $1.13\times10^{11}$ & $4.64\times10^{8}$ & $57.18$ & $3.35\times10^{9}$ & $0.84$ & 0.22 \\ 
g2.42e11 & 1,282,787 & 663,987 & 219,824 & $1.61\times10^{11}$ & $1.92\times10^{9}$ & $64.32$ & $7.48\times10^{9}$ & $0.75$ & 1.08 \\ 
g2.57e11 & 1,408,427 & 714,783 & 314,763 & $1.71\times10^{11}$ & $2.81\times10^{9}$ & $64.89$ & $6.16\times10^{9}$ & $0.61$ & 1.99 \\ 
g3.06e11 & 1,077,199 & 556,033 & 139,713 & $1.36\times10^{11}$ & $1.22\times10^{9}$ & $60.28$ & $5.08\times10^{9}$ & $0.73$ & 0.56 \\ 
g3.49e11 & 1,219,383 & 693,736 & 118,777 & $1.67\times10^{11}$ & $1.00\times10^{9}$ & $64.48$ & $4.12\times10^{9}$ & $0.70$ & 0.60 \\ 
g3.71e11 & 591,465 & 334,290 & 58,670 & $8.06\times10^{10}$ & $4.99\times10^{8}$ & $51.19$ & $2.95\times10^{9}$ & $0.80$ & 0.21 \\ 
g4.90e11 & 824,643 & 454,348 & 92,880 & $1.10\times10^{11}$ & $8.16\times10^{8}$ & $56.47$ & $4.39\times10^{9}$ & $0.76$ & 0.68 \\ 
g5.02e11 & 1,478,107 & 793,862 & 56,186 & $1.97\times10^{11}$ & $4.13\times10^{8}$ & $70.73$ & $7.18\times10^{9}$ & $0.89$ & 0.23 \\ 
g5.38e11 & 1,920,057 & 1,123,330 & 121,085 & $2.75\times10^{11}$ & $3.25\times10^{9}$ & $76.13$ & $0.98\times10^{10}$ & $0.72$ & 2.24 \\ 
g5.46e11 & 808,030 & 454,991 & 84,139 & $1.10\times10^{11}$ & $7.40\times10^{8}$ & $56.17$ & $4.08\times10^{9}$ & $0.77$ & 0.54 \\ 
g5.55e11 & 814,019 & 446,562 & 81,263 & $1.08\times10^{11}$ & $6.81\times10^{8}$ & $55.99$ & $3.82\times10^{9}$ & $0.74$ & 0.74 \\ 
g7.55e11 & 345,668 & 171,936 & 53,240 & $3.38\times10^{11}$ & $3.73\times10^{9}$ & $81.78$ & $1.71\times10^{10}$ & $0.76$ & 2.09 \\ 
g7.66e11 & 357,279 & 180,174 & 31,052 & $3.60\times10^{11}$ & $1.35\times10^{9}$ & $83.61$ & $1.17\times10^{10}$ & $0.78$ & 1.93 \\ 
g8.13e11 & 836,211 & 277,144 & 404,458 & $5.52\times10^{11}$ & $2.93\times10^{10}$ & $95.69$ & $1.36\times10^{10}$ & $0.27$ & 17.54 \\ 
g8.26e11 & 662,237 & 226,691 & 307,793 & $4.49\times10^{11}$ & $2.20\times10^{10}$ & $89.60$ & $1.47\times10^{10}$ & $0.35$ & 10.22 \\ 
g1.12e12 & 766,034 & 294,512 & 267,954 & $5.87\times10^{11}$ & $1.96\times10^{10}$ & $97.76$ & $1.95\times10^{10}$ & $0.42$ & 17.64 \\ 
g1.92e12 & 2,081,091 & 752,770 & 906,338 & $1.49\times10^{12}$ & $6.46\times10^{10}$ & $133.25$ & $2.24\times10^{10}$ & $0.22$ & 38.21 \\ 
g2.79e12 & 2,207,915 & 806,995 & 853,758 & $1.62\times10^{12}$ & $6.10\times10^{10}$ & $137.89$ & $4.76\times10^{10}$ & $0.37$ & 40.42 \\ 
\hline
\end{tabular}
\end{center}
\end{table*}

The presence of bright clumps in the light distribution of high redshift ($z\sim0.5-3$) star forming
galaxies has attracted a lot of attention both from an observational
and a theoretical point of view. Observations have  shown that  disc galaxies in  the early  universe
have higher gas-fractions \citep{Daddi2010,Tacconi2010,Tacconi2013,Genzel2015}, star formation rates  \citep{Genzel2006,Forster2006,Genzel2008} velocity dispersions \citep{Elmegreen2005,Forster2006} and a clumpy
morphology \citep{Genzel2008,Forster2011,Guo2015} compared to their counterparts at $z=0$.

Observationally these clumps  are  mostly  identified in  the
rest-frame UV,  optical or  H$_{\alpha}$ maps
 \citep{Elmegreen2007,  Elmegreen2009,  Genzel2011,
  Forster2011,  Guo2012,   Wuyts2012,  Tadaki2014,  Murata2014}, 
 although  sometimes they are  also   found  in  the   CO  line  emission  of   lensed  galaxies
 \citep{Jones2010, Swinbank2010}.
They have sizes of  about $\sim1$kpc
with  inferred  masses between  $10^7\Msun-10^9$M$_{\odot}$.
The  overall  fraction  of star-forming  galaxies  with  observed  clumps shows  significant evolution  with
redshift  \citep{Guo2015,  Shibuya2016} and increases up to $\sim$50\%. However, the number of galaxies with
observed clumps shows a large uncertainty among different measurements.

\cite{Wuyts2012} found that  the clumpy  fraction depends  sensitively on  the selected
wavelength  to  define clumps.  They  find  that the  clumpy  fraction
decreases from $\sim$75\% for clumps  selected in 2800 \AA\, images to
$\sim$40\% in the V-band.
This suggests a connection between the presence of
luminous clumps and recent star formation episodes,
as also indicated  by the observation that some clumps are  seen to be  the launching
spots of outflows  from the  disc \citep{Genzel2011}.
However, carefully constructed stellar mass  maps of galaxies show no  signs of prominent
clumps \citep{Wuyts2012}.

\begin{figure*}
\includegraphics[width=\textwidth]{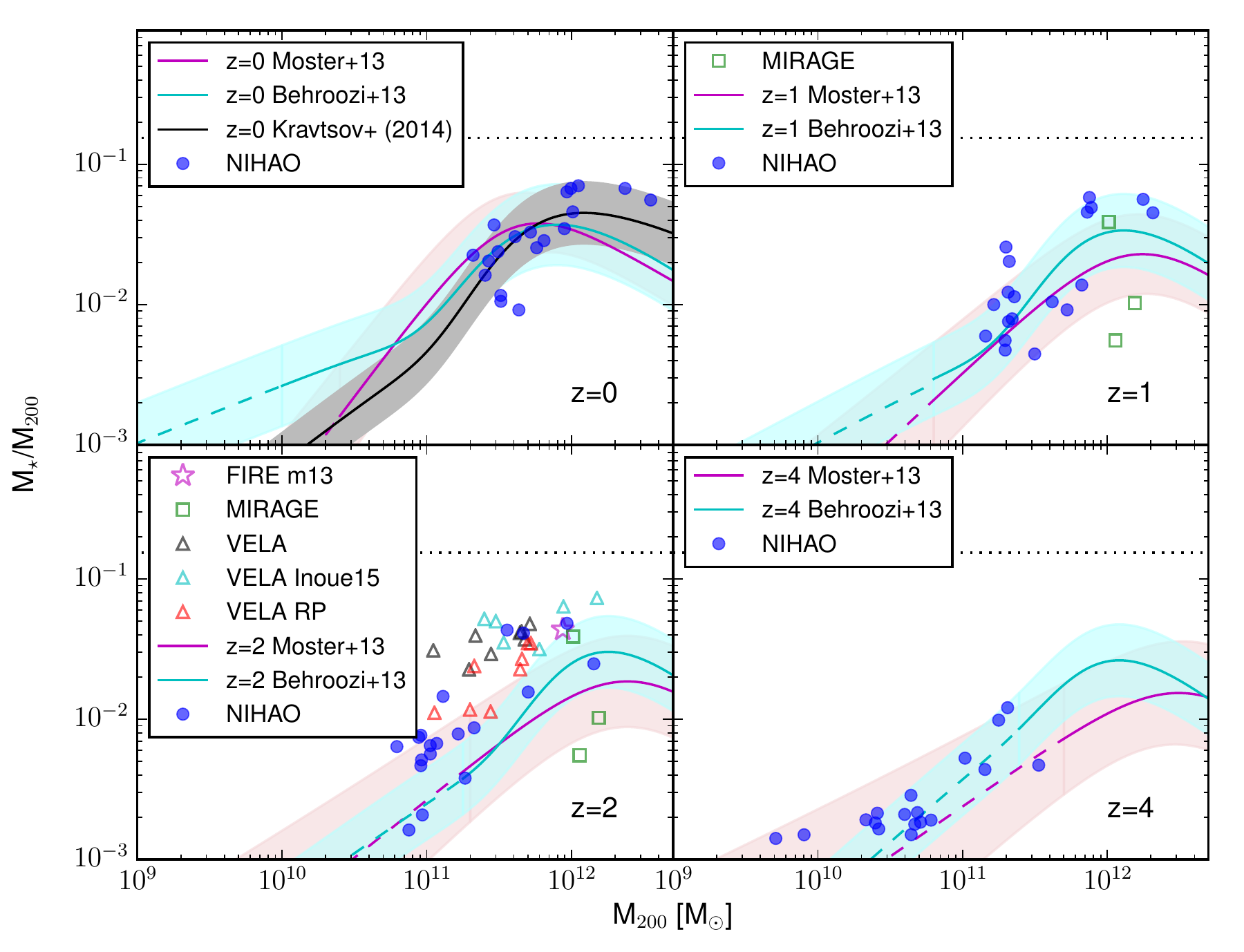}
\vspace{-.5cm}
\caption{Evolution of the Stellar mass vs. halo mass relation for the selection of hosts. Redshift z = 0 (top left), z = 1 (top right), z = 2 (bottom left), and z = 4 (bottom right). For all redshifts shown the simulations agree well with constraints from halo abundance matching \citep{Moster2013,Behroozi2013,Kravtsov2014}. Where the relations are extrapolated into mass scales without observational constraints the lines are shown in dashed line style.}
\label{fig:msmv}
\end{figure*}

Several theoretical  studies have focused  on an explanation  for the
formation  of  these  giant star-forming  clumps  either  analytically
\citep{Dekel2009a},    in    isolated    disc    galaxy    simulations
\citep{Bournaud2007,     Bournaud2008,    Bournaud2009,     Inoue2014,
  Bournaud2014,   Tamburello2015,   Mayer2016}  or   in   cosmological
simulations  of  galaxy formation  \citep{Ceverino2010,  Ceverino2012,
  Genel2012,       Hopkins2012,        Mandelker2014,       Moody2014,
  Mandelker2016}.

While some clumpy  galaxies  might be  the  result of  an
ongoing merger \citep{Somerville2001}, their overall high fraction  can not
be explained  by the  expected merger  rate at these redshifts
\citep{Dekel2009,   Stewart2009,    Hopkins2010, Hopkins2012}.
Another  explanation  for the origin of clumps, given the
high  gas  fractions  and  high  gas surface densities  in  $z>0$  galaxies,
is disc fragmentation via  gravitational (disc)
instabilities \citep{Toomre1964}.  Indeed, clumps are  mainly observed to
reside  in gravitationally  unstable  regions \citep{Genzel2011}.  The
formation scenario  invoked for  clumps is  the same as  for local
giant  molecular  clouds: local  collapse  and  fragmentation
happens in regions where the self-gravity of gas and stars is stronger
than the internal support by  pressure and turbulent motions.

As already mentioned, several groups have studied the dynamics and stability of discs with the
aid of high resolution numerical simulations 
\citep{Noguchi1998, Noguchi1999,
  Dekel2009a,  Agertz2009, Cacciato2012, Ceverino2015,  Genel2012,    Inoue2012,   Perez2013, Tamburello2015}.
In most  of these  studies discs  do  fragment and  break  up  into  large clumps,
however  there are two different possible scenarios for the fate of the clumps
and their subsequent impact on the global evolution of the host galaxy.
If clump collapse is very efficient, it could lead to the formation of
gravitationally self-bound,  long-lived giant  star clusters.
These clusters will then lose angular momentum  within the
disc  via gravitational  torques  and dynamical  friction and  migrate
inwards  on  timescales  of  about  $10^8$yr to  build  up  the  central bulge
\citep[e.g.][]{Bournaud2014}.
This  picture  seems to be  supported  by  observed
color gradients  of the clumps within  the disc. Clumps closer  to the
center  of  a  galaxy show  redder  colors  \citep{Forster2011,
  Guo2012, Shibuya2016}  although this  trend appears  to be  weak and
might  be caused  by underlying  evolved structures  like e.g.  bulges
\citep{Dokkum2010, Patel2013, Morishita2015, Nelson2016}.

Conversely if gas cooling is suppressed, as for example
in the presence of substantial stellar feedback, the resulting
stellar clumps will not be self-bound and will then be quickly dispersed within the disc before experiencing any
drag to the center.

Recent numerical work \citep{Mayer2016,Oklopcic2016}  has somehow suggested
that  when  robust feedback  (needed   to  reproduce
observed galaxy properties at  redshift $z=0$)   is
invoked, clumps  are either  short lived transient  features or  do not
form at all. Thus a treatment of feedback, that is able to prevent
overcooling seems to be crucial to understand the origin of
clumpy galaxies at high redshift.
On the other hand these studies were either based on a single
simulated galaxy \citep{Oklopcic2016} or they were lacking
cosmological gas inflow  \citep{Mayer2016} and hence
more work on the simulation side is needed.

In this work we revise the issue of the formation and evolution
of luminous  clumps in cosmological simulations of disc galaxies at $z=1-3$
using the NIHAO simulation suite. 
The NIHAO (Numerical Investigations of Hundred Astrophysical Objects)\footnote{Nihao is the
  chinese word for {\it hello}.} project is a suite of one hundred high resolution hydrodynamical cosmological
simulations. We first compare the ``clumpiness'' of our galaxies with real 
galaxies using an observational motivated clump  selection procedure
based on the UV luminosity of our objects.
For this purpose we have post processed all our galaxies with the radiative transfer code {\texttt{GRASIL-3D}}.
\citep{Dominguez2014}.
Subsequently we look at the evolution of these luminous clumps and their relation
with the underlying stellar mass distribution to assess their final fate
and their overall impact on galaxy evolution.

\section{Simulations} \label{sec:simulation}

This work is based on the NIHAO simulation suite \citep{Wang2015} which
contains  nearly 100 zoomed cosmological simulations of galaxy formation across a wide range
in stellar mass from $10^5  < M_{\star} / M_{\odot}  < 10^{11}$.
One peculiarity of the NIHAO suite is that the numerical resolution, i.e. the number of elements (particles)
representing each galaxy, is kept roughly constant over the whole mass range
with of the order of one million particles (gas+stars+dm) in each simulated galaxy.

The NIHAO galaxies have been run using cosmological parameters from the \cite{Planck}, 
namely: \OmegaM=0.3175,   \OmegaL=0.6825,   \Omegab=0.049,  H${_0}$   =   67.1
\kms\Mpc$^{-1}$, \sig8  = 0.8344.
For the sub-set of simulations used here the mass resolution is   either   $m_{\rm    dark}=1.735\times10^6   \Msun$   or
$m_{\rm  dark}=2.169\times10^5 \Msun$ for dark matter particles  and $m_{\rm  gas}=3.166\times10^5  \Msun$ or
$m_{\rm gas}=3.958\times10^4  \Msun$
for   the  gas   particles.   The  corresponding force  softenings   are
$\epsilon_{\rm dark}=931.4$ pc  and $\epsilon_{\rm dark}=465.7$ pc
for the dark matter particles  and $\epsilon_{\rm gas}=397.9$ pc and
$199.0$ pc for the gas and star particles, which is the spatial scale more relevant for this study.
 However, the  smoothing length of
the  gas  particles  can  be  much  smaller,
e.g.  as  low  as  $h_{\rm  smooth}\sim 45$ pc.
See Table \ref{tab:sims} for a full list of the our galaxies parameters.

NIHAO galaxies have been proven to match remarkably well many of the properties
of observed galaxies, as for example results from abundance matching
\citep{Wang2015}, metals distribution in the Circum Galactic Medium \citep{Gutcke2016},
the local velocity function \citep{Maccio2016} and the properties of stellar
and gaseous discs \citep{Obreja2016, Dutton2016b}.
Overall NIHAO simulated galaxies are among the most realistic simulations
run in a cosmological context. This means that the NIHAO sample is a
perfect test bed to study the occurrence and properties of clumps
in high redshift galaxies.

\subsection{Hydrodynamics}

The NIHAO simulations were run with a  modified version of the  smoothed
particle hydrodynamics  (SPH) solver {\texttt{GASOLINE}}  \citep{Wadsley2004}
  with  substantial  updates  made   to   the   hydrodynamics  as described   in
  \citep{Keller2014}. We will refer to this version of {\texttt{GASOLINE}} as \texttt{ESF-GASOLINE2} .

This  modified  version of  hydrodynamics  removes
spurious numerical  surface tension and improves  multi-phase mixing by
calculating $P/\rho^2$ as a geometrical  average over the particles in
the smoothing  kernel as proposed  by \cite{Ritchie2001}.
The  \cite{Saitoh2009} timestep limiter
was implemented  so that  cool particles behave  correctly when  a hot
blastwave hits them  and \texttt{ESF-GASOLINE2} uses the  Wendland C2 smoothing
kernel   \citep{Dehnen2012}  to   avoid  pairing   instabilities.  The
treatment of artificial viscosity has  been modified to use the signal
velocity as described  in \cite{Price2008} and the  number of neighbor
particles  used  in  the  calculation  of  the  smoothed  hydrodynamic
properties was increased from 32 to 50.

Cooling via hydrogen, helium,
and various  metal-lines is  included as described  in \cite{Shen2010}
and     was     calculated     using    \texttt{cloudy} 
\citep[version     07.02;][]{Ferland1998}.
These  calculations include photo  ionization and
heating from  the \cite{Haardt2005} UV background  and Compton cooling
in a temperature range from 10 to $10^9$ K.
Finally we adopted a metal diffusion algorithm between particles
as described  in \cite{Wadsley2008}.

\subsection{Star Formation and Feedback}
In the fiducial NIHAO runs gas  is eligible to form stars according to
the Kennicutt-Schmidt Law when it  satisfies a temperature and density
threshold.   The  simulations  employ  the star  formation  recipe  as
described in \cite{Stinson2006} which  is summarized below. Stars form
from  cool  (T  $<  15,  000  $K), dense  gas  ($n_{\rm  th}  >  10.3$
cm$^{-3}$).  The  threshold number  density  $n_{\rm th}$  is set  to  the
maximum density  at which gravitational instabilities  can be resolved
in  the  simulation:  n$_{\rm th}$  $=$  50$m_{\rm  gas}/\epsilon_{\rm
  gas}^3 = 10.3$ cm$^{-3}$, where $m_{\rm gas}$ denotes the gas particle
mass and $\epsilon_{\rm gas}$ the gravitational  softening of the gas. The value of 50 denotes the number of neighboring particles.
 The gas fulfilling  these requirements will  then be converted  into stars
according to the following equation:
\begin{equation}
\frac{\Delta  M_{\rm star}}{\Delta t}=c_{\rm star}\frac{M_{\rm  gas}}{t_{\rm
    dyn}}
\end{equation}
where  $\Delta M_{\rm star}$  is the  mass of  the star  particle formed,
$M_{\rm  gas}$ the  gas particle's  mass, $\Delta  t$ is  the timestep
between star formation  events (here: $8 \times 10^5$  yr) and $t_{\rm dyn}$
is  the gas particles dynamical  time. $c_{\rm star}$
is the star  formation efficiency, i.e. the fraction of  gas that will
be converted into stars during the  time $t_{\rm dyn}$ and is taken to
be $c_{\rm star}=0.1$.

Stellar  feedback is implemented in  two modes as
described  in  \cite{Stinson2013}. The  first  mode  accounts for  the
energy  input from  stellar  winds and  photo ionization from  luminous
young  stars.   This  pre-SN   feedback, that was  dubbed Early Stellar Feedback
(ESF) in \cite{Stinson2013}, 
happens  before any  supernovae explode  and consists  of 10\%  of the
total  stellar flux,  $2 \times  10^{50}$  erg of  thermal energy  per
$M_{\odot}$ of the entire stellar population. The efficiency parameter
for the  coupling of the  energy input  is set to  $\epsilon_{\rm ESF}
=13\%$. And for the pre-SN feedback, unlike the SN feedback, the radiative
cooling is left on.

The second mode accounts for the energy input via
supernovae and starts 4 Myr after  the formation of the star particle,
and it is implemented using the blastwave  formalism described  in \cite{Stinson2006}.
In this approach super novae input energy (i.e. thermal feedback) into  the interstellar  gas  surrounding the
region were they formed, but since the  gas
receiving the energy  is dense, it would quickly be  radiated away due
to  its efficient  cooling. For  this reason,  cooling is  delayed for
particles inside  the blast region for  $\sim$30 Myr. See \cite{Stinson2013} for
an extended feedback parameter search.

\begin{figure*}
\includegraphics[width=\textwidth]{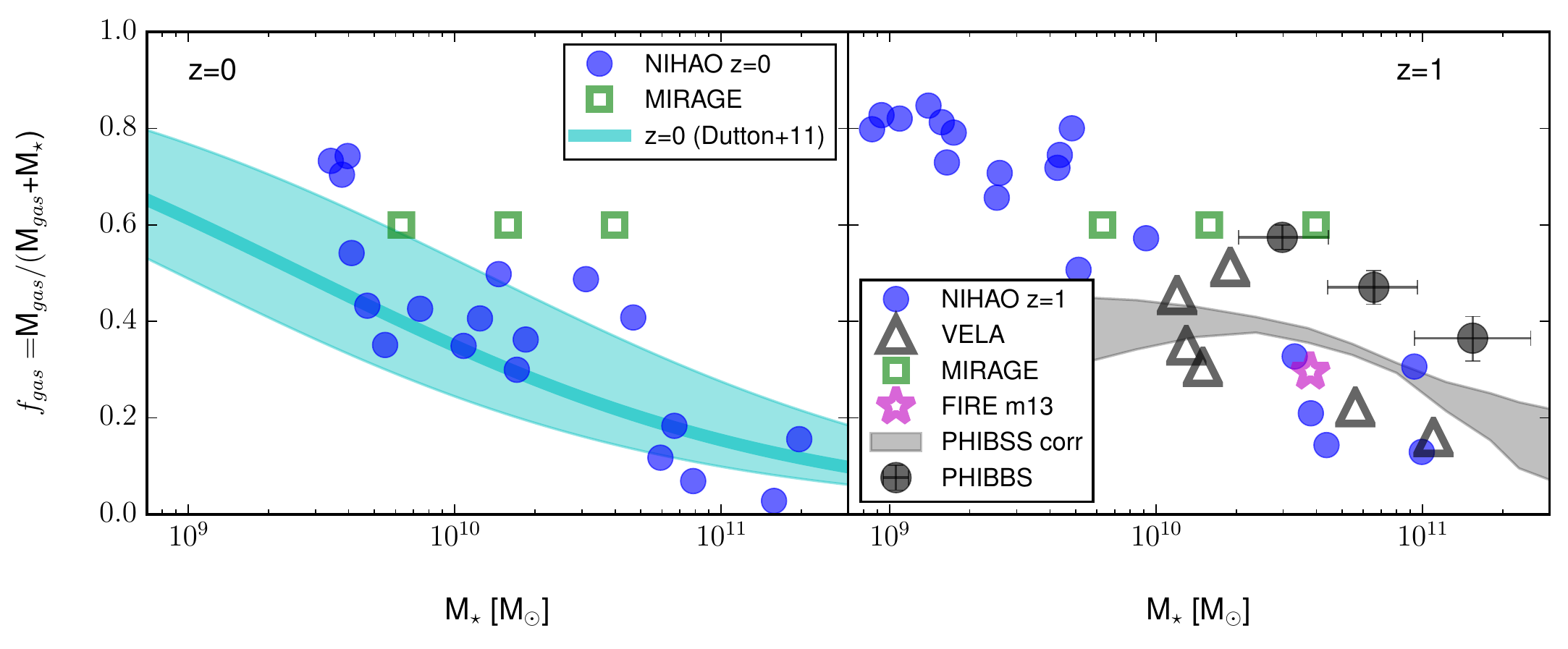}
\vspace{-.35cm}
\caption{The galaxy gas fraction as a function of stellar mass at redshift $z=0$ (left panel) and $z=1$ (right panel). The selected NIHAO galaxies are shown with blue points and the relation from \protect\cite{Dutton2011} for redshift 0 galaxies is shown with a cyan line with a shaded region indicating the scatter. At redshift $\sim1.5$ the PHIBBS survey data \protect\citep{Tacconi2013} is shown as black dots. The incompleteness corrected relation from the PHIBBS survey is shown as the grey shaded area. }
\label{fig:gasfrac}
\end{figure*}

\section{The Galaxy sample}

In order to match the stellar range of the observed galaxies at high redshift
we have selected from the NIHAO sample all the galaxies with stellar masses
larger than $M_{\rm star}>10^{9}M_{\odot}$ at redshift $z=1.5$, obtaining
a final sample of 19 galaxies. From these galaxies we use all the snapshots in the redshift
range $3>z>0.25$ where the stellar mass of the galaxy is larger than the above threshold mass.
 This selection criteria leaves us with 203 snapshots in the redshift range $3>z>1$, 155 in the redshift range
 $1>z>0.5$ and 130 in the redshift range $0.5>z>0.25$.

For this work the  virial mass, $M_{200}$, of each halo  is defined as the
mass of all particles within a  sphere containing $\Delta$ = 200 times
the  cosmic  critical matter  density,  \rhocrit.  The virial  radius,
$R_{200}$, is defined accordingly as the radius of this sphere. The haloes
in the zoom-in simulations were identified using the MPI+OpenMP hybrid
halo finder  \texttt{AHF2} \citep{Knollmann2009,  Gill2004}.
The  stellar $M_{\rm star}$  and the gas, $M_{\rm   gas}$ masses are measured
within a sphere of radius, $r_{\rm gal}$ $\equiv$ 0.2$R_{200}$.
The  star formation rate,  SFR, is  measured as the  mass of
stars formed inside  $r_{\rm gal}$ over the preceding Gyr  and the gas
fraction $f_{\rm gas}$  is defined as the fraction of  cold gas (T $<$
$3\times10^4$ K) over the total  baryonic  mass within $r_{\rm  gal}$.
The main  parameters of the 19 NIHAO galaxies are listed in Table 1.

As already mentioned before the strength of the (stellar) feedback has
a strong impact in promoting or suppressing the formation of stellar
clumps. The efficiency of the feedback can be tested by looking
at how realistic the properties of simulated galaxies are w.r.t. observed
ones.

\subsection{Stellar mass-halo mass relation}

Fig. \ref{fig:msmv} shows the evolution of the stellar mass
vs.  halo mass  relation since  redshift z  = 4  (a look  back time  of
$\sim$12 Gyr); NIHAO simulations (blue circles) show  a very
good agreement with the abundance matching   relations   from \cite{Behroozi2013} and
\cite{Moster2013}.
At redshift 2 and 1 some of the higher mass galaxies show about
a factor of 2 to many stars  but this is consistent with e.g. the FIRE
simulation     \cite{Hopkins2014}.
Some of  this discrepancy w.r.t abundance matching results
might be  due to systematic uncertainties  in the form of  the stellar
Initial  Mass  Function (IMF, e.g. \cite{Conroy2012,
  Dutton2013,  Dutton2013b}).
Thus  for  massive galaxies  ($M_{\star}
\gsim 10^{11}  \Msun$) the stellar  masses may be underestimated  by a
factor  of  ~2  when  assuming  a Milky  Way  IMF.

Comparison     to    the    VELA    galaxies
\citep{Ceverino2015,  Inoue2016,  Moody2014}   (colored  triangles  in
Fig. \ref{fig:msmv}) shows these galaxies tend to substantially overproduce
stellar masses  at redshift 2.
Inclusion of radiation pressure  as an
additional source of  feedback in these galaxies  (VELARP) brings them
in  better agreement  with the  abundance matching  relation.
This  in comparison  to the  results from  the FIRE  simulation and  from NIHAO
indicates that inclusion  of some sort of feedback  prior to supernova
(Ceverino:  radiation pressure;  Hopkins: radiation  pressure, stellar
winds, and photoionization; NIHAO:  strong photoionization included as
thermal  energy) is  needed to  reproduce the  stellar mass-halo  mass
relation at high redshifts.

It is worth noticing that the VELA galaxies were not run down to  redshift 0
so there is  no information whether these simulations
do or do not provide realistic present day galaxies.
For reference   we    also   include    the   isolated    simulations   by
\cite{Perret2014} taken from the MIRAGE sample.

\subsection{Gas fractions}

Together with the total stellar mass, another important  quantity in determining
the  stability of a disc in a galaxy is the fraction  of cold  gas.

In  Fig. \ref{fig:gasfrac}  we  show the  gas
fraction defined as  $f_{\rm gas}=M_{\rm gas}/(M_{\star}+M_{\rm gas})$
as a function  of stellar mass compared to observations  at redshift 0
from  \cite{Dutton2011} (left panel)  and  observations   from  the  PHIBBS  survey
\citep{Tacconi2013} for galaxies at redshift 1-1.5 (right panel). The selected NIHAO
galaxies  follow  nicely the  trend observed  by \cite{Dutton2011}.

The discrepancy at these lower  masses might be due to different
measurement methods  in the simulations  and the observations.  In the
simulations  we  use all  the  gas  with  a temperature  smaller  than
$3\times10^4$ K in a sphere of radius $r_{\rm gal}=0.2R_{200}$ while the
observations measure atomic and molecular gas with a correction factor for helium
\citep{Dutton2011}.

As  was  shown  in  \cite{Stinson2015}  a  simple
temperature cut overestimates the  amount of neutral gas, especially
at lower masses.  The galaxies are also broadly in  agreement with the
incompleteness  corrected measurements  from the  PHIBBS survey  (gray
band) for galaxies  in the redshift range $z=1-1.5$.  For reference we
also show the gas fractions  of the VELA galaxies \citep{Ceverino2015,
  Inoue2016, Moody2014}, the FIRE  galaxy from \cite{Oklopcic2016} and
the MIRAGE  sample \citep{Perret2014}.  The VELA  sample and  the FIRE
galaxy are  well in  agreement with  the NIHAO sample  as well  as the
observed gas fractions. The two  higher mass MIRAGE galaxies show very
high  gas fractions  of  60\%  in slight tension with the completeness
corrected  PHIBBS observations. Although galaxies with such high gas fractions are observed
they are likely not typical.

\section{Radiative Transfer}
\label{sec:RT}

\begin{figure*}
\includegraphics[width=\textwidth]{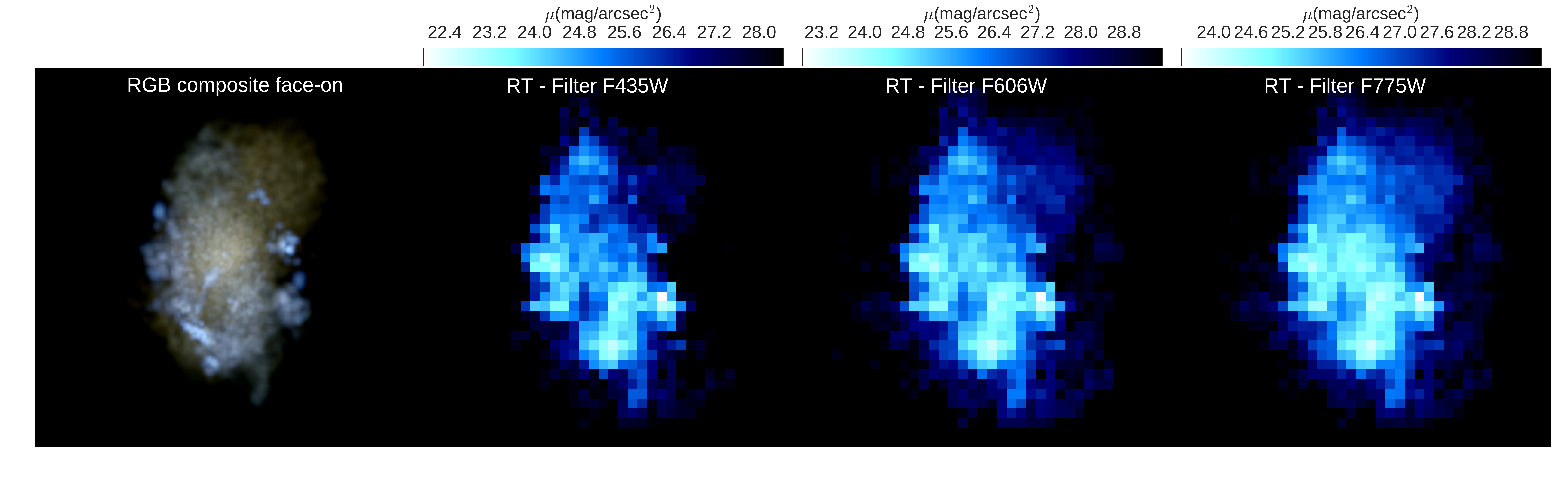}
\vspace{-.35cm}
\caption{RGB image (left panel) and radiative transfer (RT) maps in different filters. The second panel shows the map of the F435W filter, the third panel shows the F606W image and the most right panel shows the F775W filter. Image sizes are 20 x 20 kpc.}
\label{fig:GRASIL}
\end{figure*}

In order to compare the  fraction of  galaxies with clumps  in the simulations and in the
CANDELS galaxy sample we post-process our simulations with the radiative transfer code {\tt GRASIL-3D} \citep{Dominguez2014}.
The post-processing step insures we account for the effects of dust attenuation and cosmological redshift,
thus allowing for a meaningful comparison between simulations and observations. 

{\tt GRASIL-3D} is a three-dimensional radiative transfer code designed to be used with the outputs of hydrodynamical simulations. 
The code solves the radiative transfer equation for dusty media on a regular grid. The treatment of dust is based on the 
formalism of the {\tt GRASIL} model \citep{Silva1998, Granato2000}, which has been successfully used with 
semi-analytical models of galaxy formation. The key feature of this dust model is that it does a detailed non-equilibrium 
calculation for polycyclic aromatic hydrocarbons molecules and dust grains smaller than 150\AA, thus allowing for a 
proper description of the cirrus emission in mid-infrared \citep{Guhathakurta1989}.  

An important point to remember is that any RT post-processing of simulated galaxies introduces a few more sub-grid 
parameters on top of those already included in the hydrodynamic codes. In the case of {\tt GRASIL-3D}, these
parameters are particularly related to the properties of the molecular clouds (MCs). Since short-lived massive stars are
spatially associated with MCs, it is well established that much of the dust reprocessing of stellar light occurs inside MCs. 
In the case of most cosmological simulations these small scales are not resolved, and as such, 
dust reprocessing has to be modeled. In {\tt GRASIL-3D}, the interstellar medium is split into MCs and diffuse cirrus, 
by assuming the densities of unit hydrodynamical gas masses to follow a lognormal probability distribution function 
with a mean equal to the local gas density. The dispersion of the distribution is a free parameter. 
In this manner, the fraction of the gas mass above a certain density threshold gives the MC contribution, while the rest 
is considered cirrus. Thus, the first two parameters {\tt GRASIL-3D} needs are the MC density threshold and the 
dispersion of the density distribution function. 

In {\tt GRASIL-3D} the dust reprocessing of stellar populations is age-dependent, similar to the implementation in {\tt GRASIL}
which was the first model to take age into account. Practically, stars younger than a certain age, t$_{\rm 0}$, 
radiate all their energy inside MCs, while stars older than 2t$_{\rm 0}$ have already dispersed their MC cocoons.
In the intermediate age regime, the fraction of energy dumped inside the MC is a decreasing function of stellar age. 
The last parameter the code needs is the spatial extent of molecular clouds.
Once these sub-grid parameters are set, {\tt GRASIL-3D} solves the RT equation by treating separately the light reprocessing 
in the dense and diffuse interstellar medium. 

Finally, the stellar particles luminosities are computed according to the simple stellar population models of 
\cite{Bruzual2003} (which is the only option available in the current version of {\tt GRASIL-3D }) 
assuming a Chabrier IMF \citep{Chabrier2003}. 

The code has already been used to study the star formation main sequence of simulated galaxies \citep{Obreja2014}, 
the properties of high redshift clusters and proto-clusters in sub-mm and IR \citep{Granato2015}, and
the correlations between IR fluxes of z=0 simulated galaxies and their baryonic content \citep{Goz2016}. 
These studies have shown that {\tt GRASIL-3D} reproduces observables (e.g. broad band fluxes, spectral energy distributions) 
when used with realistic galaxy simulations.

In order to run {\tt GRASIL-3D} on the simulated NIHAO galaxies, we need to specify the four parameters discussed above. 
We chose the fiducial values for disc galaxies of \cite{Dominguez2014}: 
14 pc for the molecular cloud radius, 5 Myr for the molecular cloud destruction time-scale,
3.3$\times$10$^{\rm 9}$M$_{\odot}$kpc$^{\rm -3}$ for the molecular clouds threshold density, 
and a dispersion of the log-normal gas density probability distribution function of 3. 
We do not perform any {\tt GRASIL-3D} parameter study in this work, 
given both the high computational cost implied and the fact 
that the above mentioned values have been shown to reproduce normal star forming galaxies \citep[e.g.][]{Silva1998, Goz2016}.
Finally, we chose  pixel sizes  corresponding to the  resolution of the Hubble
Space Telescope (HST) at the  given redshift of the snapshots (0.06''). 

After running {\tt GRASIL-3D} on all the snapshots of our galaxy sample in the redshift range $\sim$0.25-3,
we apply the same  HST filter  selection with the  same magnitude
cuts and  surface brightness limits  as chosen in  \cite{Guo2015}.
For snapshots in  the redshift range  $3>z>2$ we select the  F775W filter,
for $2>z>1$  we select the  F606W filter and  for $z<1$ we  select the F435W  filter   to  detect  clumps.  
The  outcome of  the radiative  transfer calculations results in mock  observations of our simulations  closely matching the
ones of CANDELS galaxies used by \cite{Guo2015}. 
A selection of images for the galaxy g7.55e11 at redshift $z\sim1.3$ in all three filters used are shown in  Fig. \ref{fig:GRASIL}  in comparison to
an RGB map of the stellar luminosity (left panel). For the RGB map we calculate the stellar luminosity of a star particle given 
its age and metallicity in three different bands (see next section for a more detailed description). 
The radiative transfer images show the luminosity maps of the same galaxy in the three filters used for the analysis (from left to right: F435W, F606W, F775W). 
As described above, for the given snapshot time of $z\sim1.3$ we would use for the clump analysis the F606W image, 
while for higher redshifts we would use the F435W filter and for lower redshifts the F775W one.

\section{Clump Detection}
\label{sec:detection}

In this study we perform an  observationally motivated  clump
selection, while other theoretical  works on  this subject focused
on identifying  clumps as regions  of high  surface density of  gas or
stars \citep{Genel2012,Ceverino2015,Tamburello2015,Mayer2016,Inoue2016,Oklopcic2016}

Observed  clumps  are  mostly detected  as  UV  bright  or
$H_{\alpha}$ bright clumps and we thus decided to select  clumps  in the
luminosity maps  of  our galaxies  with and without radiative transfer post-processing.
We will refer to clumps detected in non dust-attenuated images as {\it intrinsic} clumps,
while we will call clumps in the RT-processed images {\it observed } clumps.
This allows us to do a proper comparison with both observations and previous theoretical studies.

\subsection{Intrinsic clump selection}
\label{ssec:clumpselection}

For every galaxy  and every snapshot in the redshift  range $z=0.25-3$ we
create UV-light images by calculating  the UV luminosity of every star
particle given its  age, metallicity and its IMF  under the assumption
that these  particles represent simple stellar  populations (SSPs). We
use the {\sc{pynbody}}-package\footnote{https://pynbody.github.io/pynbody/} to
perform  these  calculations. This  package  includes  a grid  of  SSP
luminosities for  different stellar ages and  metallicities in several
bandpasses.  The  grids are  calculated  using  Padova Simple  stellar
populations from Girardi\footnote{http://stev.oapd.inaf.it/cgi-bin/cmd}
\citep{Marigo2008, Girardi2010}.

Our  clump   finding  procedure  is   similar  to  the  one   used  by
\cite{Oklopcic2016} and we will briefly  describe it here.
For all snapshots we  first rotate the  galaxies face-on, using
the total  angular momentum  of the stars within a sphere  of 10 \kpc around
the center of the halo. 
We then  select  all star particles in a cylinder of radius  10 \kpc and height of 6 \kpc centered
on the galaxy and then construct the luminosity
maps by  binning the  particle positions  onto a  2d-grid of  bin size 100 pc.

\begin{figure*}
\includegraphics[width=\textwidth]{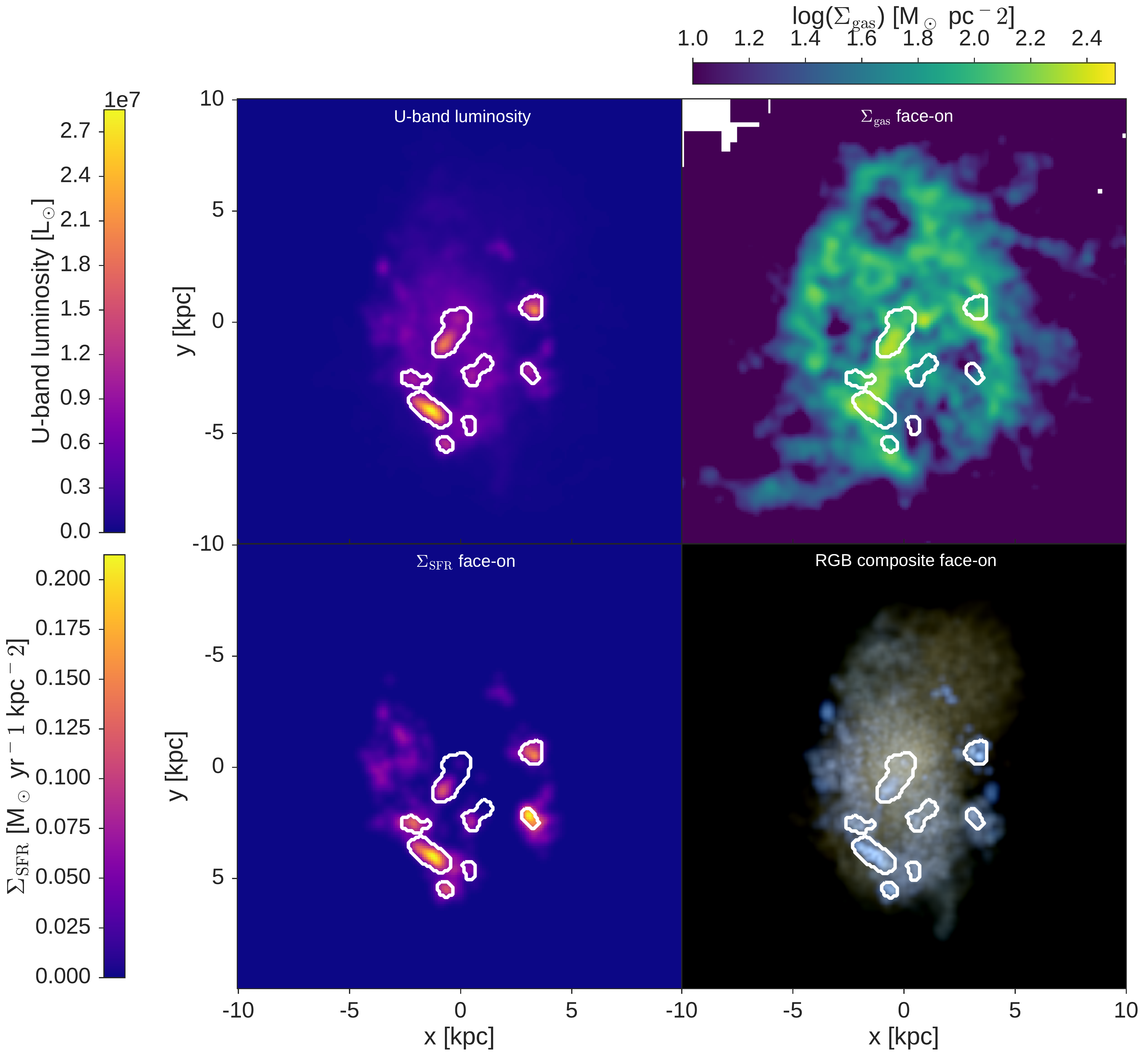}
\vspace{-.35cm}
\caption{Face-on maps  of intrinsic u-band luminosity (upper  left), SFR surface
  density (lower left),  cold gas surface density  (upper right),
  and RGB  stellar  composite  (lower  right). The
  white contours show the clumps selected in the intrinsic u-band luminosity map.}
\label{fig:surf}
\end{figure*}

\begin{figure*}
\includegraphics[width=\textwidth]{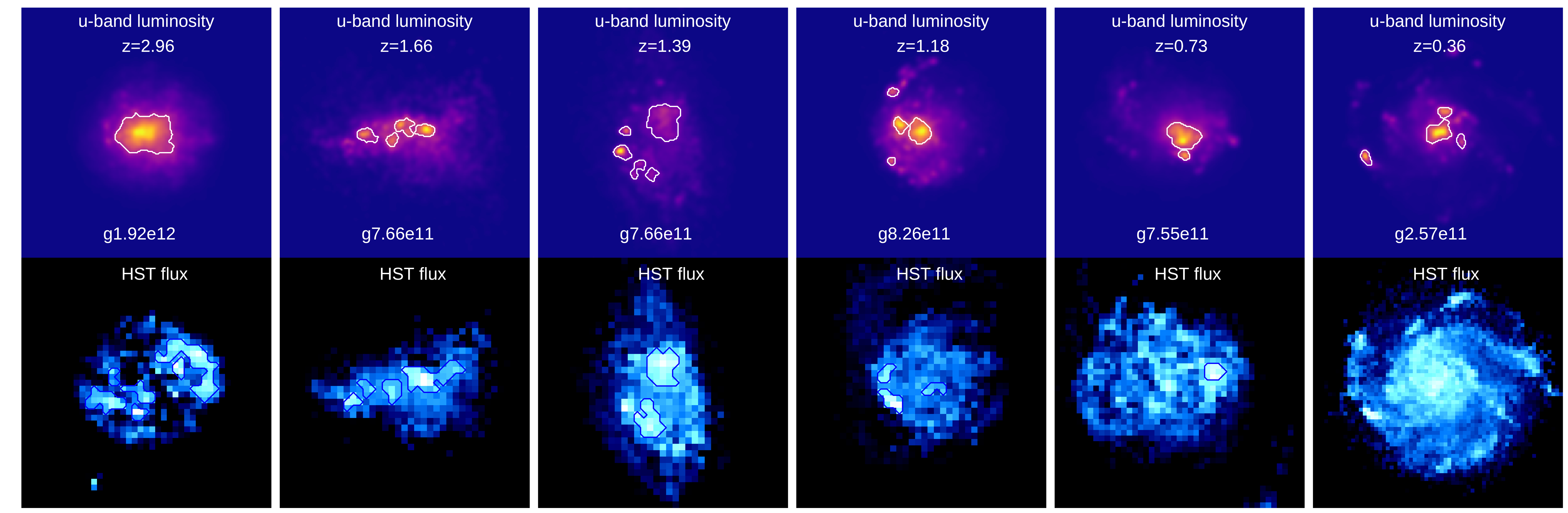}
\vspace{-.35cm}
\caption{Comparison between clumps in the rest frame u-band luminosity
  maps (upper panels)   and clumps in the radiative transfer (RT) images (lower panels). For
  the RT images the galaxies were redshifted to the proper redshift
  of  the snapshot  and  observed  in the  same  filter  bands as  the
  observations   by   \protect\cite{Guo2015}   accounting   for   dust
  absorption, scattering and cosmological redshift. For galaxies in
  the  redshift  range  $3<z<2$  we selected  the  F775W  filter,  for
  $2<z<1$we selected F606W and for $z<1$ we selected F435W, where the redshift of each panel is indicated in the u-band maps.}
\label{fig:surf_GRASIL}
\end{figure*}

We sum up all  luminosities of the star
particles in one  cell to get its total luminosity.
After that  we smooth the luminosity map by  convolving it with a
Gaussian filter of 200pc standard deviation (FWHM $\sim$470pc) to
account for the particle's softening. We checked that taking
the softening into account before binning the data  onto the grid does
not make a significant difference in terms of clump detection.

Using the procedure described above
we  can also  create maps  of different  quantities, e.g.  the surface
density maps of  gas can be calculated  by summing up all  the mass in
one  bin and  then dividing  the  result by  the surface  area of  the
bin. Figure  \ref{fig:surf} shows the u-band luminosity  map (upper left),
the SFR surface density map (lower left) with the SFR calculated as the stellar mass formed over the last 500 Myr, surface density of cold gas (upper right)  and a RGB stellar
composite image of the stars (lower right), where RGB stands for the colors used to render the stars (old stars are rendered with a Red color, intermediate aged stars with Green and young stars Blue).
As expected, intrinsic clumps  found in the  u-band luminosity agree  very well
with clumps in the SFR surface density. Once we
created a map of a given  quantity we further calculate the mean value
of all  non-empty bins and  the according standard deviation.
We then use the Python package
astrodendro\footnote{http://www.dendrograms.org/}        to       find
over-densities  in  the  maps.   The  astrodendro  package  calculates
hierarchical  trees  of  structures   so  called  dendrograms  in  the
maps. There  are three parameters  needed by the package  to calculate
the tree. i) A threshold value to define the clump which we set to
three standard deviations above the mean value (we checked that our results are robust if we alter this value, see fig. \ref{fig:sensitivity} in the Appendix),
 ii) The minimum difference between two close structures to count them as separate clumps; we set
this to 10\% following \cite{Oklopcic2016}. iii) The minimum number
of pixels within a clump which is  set to be 30 pixels resulting in an
effective radius  of $\sim$300 pc which  is consistent with the gravitational softening 
of the gas and star particles of our simulations.

An example  of the  outcome of this clump finding algorithm is shown in
Figure \ref{fig:surf} where u-band selected clumps are over-plotted on all four face-on images with white contours. 

The astrodendro package  already comes with tools to  measure the size
(surface area) of each clump in  the plane of the disc (x-y-plane). Given the surface area, $A$, of
each clump  we can calculate  an effective radius  of the clump,  $R^2 =A/\pi$.
We     follow    the     assumption  by \cite{Oklopcic2016} and take the extent of each clump perpendicular to
the plane of the disc as equal to $2R$ centered on the densest part of
the clump. All  stellar and gas particles falling into  this volume of
space are counted as belonging to  the clump and clump properties such
as luminosity, mass, etc. can  be calculated from these particles.  We
checked that the  enclosed mass does not depend strongly  on the exact
choice of the clump's  vertical extent, as long as it  is on the order
of the disc  scale height. Most of the snapshots  show one large clump
in  the  center of  the  galaxy  which can  be  matched  to the  bulge
component. Thus, we excluded from the search area the innermost 1 kpc
around the center of the galaxy.

\subsection{Clump selection in radiative transfer images}
\label{ssec:RTclumps}

While the above clump selection is useful to study physical properties of clumps,
a comparison to observed galaxies is difficult due to missing dust attenuation and cosmological redshift.
Therefore we look for clumps directly in the dust attenuated radiative transfer (RT) maps computed as explained in section 4.

On  the  RT  images  we adjust our clump  finder  to
better match the observational  clump selection as described in  \cite{Guo2015} (section  3).
We first calculate  the background  mean
after   applying a 3$\sigma$ clipping  and then we select clumps  as local
maxima which are at least 3$\sigma$ above the  mean. Again we checked how results change if this threshold is changed to 2 or 4$\sigma$ (see fig. \ref{fig:rad_tr_comp} in the Appendix).
Because we adjusted the pixel size of the RT images to match the resolution of HST we now set the minimum  
number of pixels to 5 per clump following \citep{Guo2015} and we
neglect clumps within  the inner 1 kpc around the  galaxy center.

Fig. \ref{fig:surf_GRASIL} shows that the RT images recover some of the
clumps found  in the  U-band images  but not  all.
Sometimes  the dust attenuation and  the cosmological redshift  dim down clumps  such that
they  fall below  the detection  threshold of  $3\sigma$ above  the mean
(e.g. fourth  panel of  Fig. \ref{fig:surf_GRASIL}).  On  the other
hand a non-clumpy  galaxy in the U-band shows clumps  in the RT
 image due to a  non-uniform gas distribution (e.g. first and fiftf panel of Fig. \ref{fig:surf_GRASIL}). This underlines the
importance of using radiative transfer images to  compare simulation and
observations quantitatively.

\section{Results}
\label{sec:results}

\subsection{The \textit{observed} NIHAO clumpy fraction}

\begin{figure*}
\includegraphics[width=\textwidth]{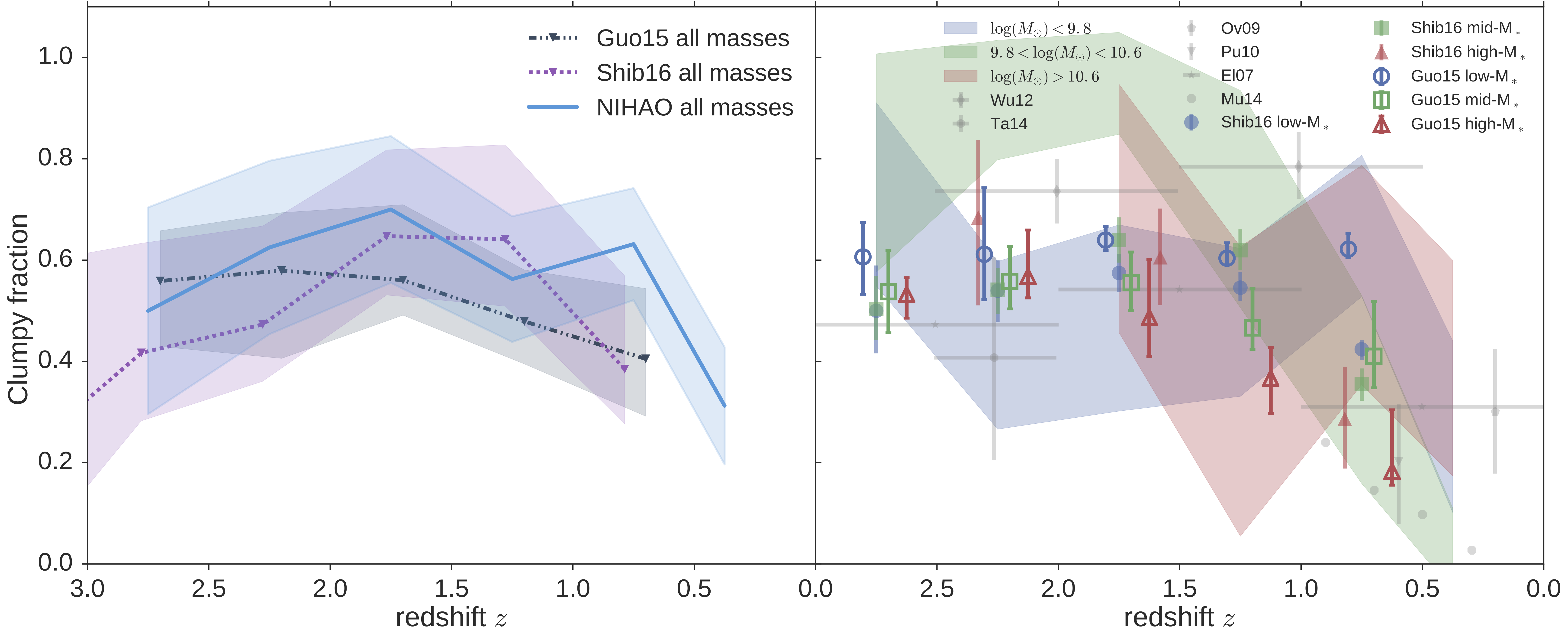}
\vspace{-.35cm}
\caption{Evolution of the fraction of galaxies with at least one \emph{observed} off-center clump. \emph{The left panel} shows the evolution of the clumpy fraction for the whole NIHAO sample (blue line) compared to observations from \protect\cite{Guo2015} (black dash-dotted line) and \protect\cite{Shibuya2016} (purple dotted line). The shaded band shows the $1\sigma$ scatter. 
\emph{The right panel} shows the evolution of the clumpy fraction splitted into different mass bins.
Colored bands show the result for our simulations, colored open symbols show the according values from \protect\cite{Guo2015} and colored filled symbols show the results from \protect\cite{Shibuya2016}. Grey open symbols show results from other observational studies: diamonds for \protect\cite{Wuyts2012}, squares are from \protect\cite{Tadaki2014}, pentagons are from \protect\cite{Overzier2009}, inverse triangles from \protect\cite{Puech2010}, stars from \protect\cite{Elmegreen2007} and dots from \protect\cite{Murata2014}.}
\label{fig:clumpy_frac}
\end{figure*}

\begin{figure}
\includegraphics[width=\columnwidth]{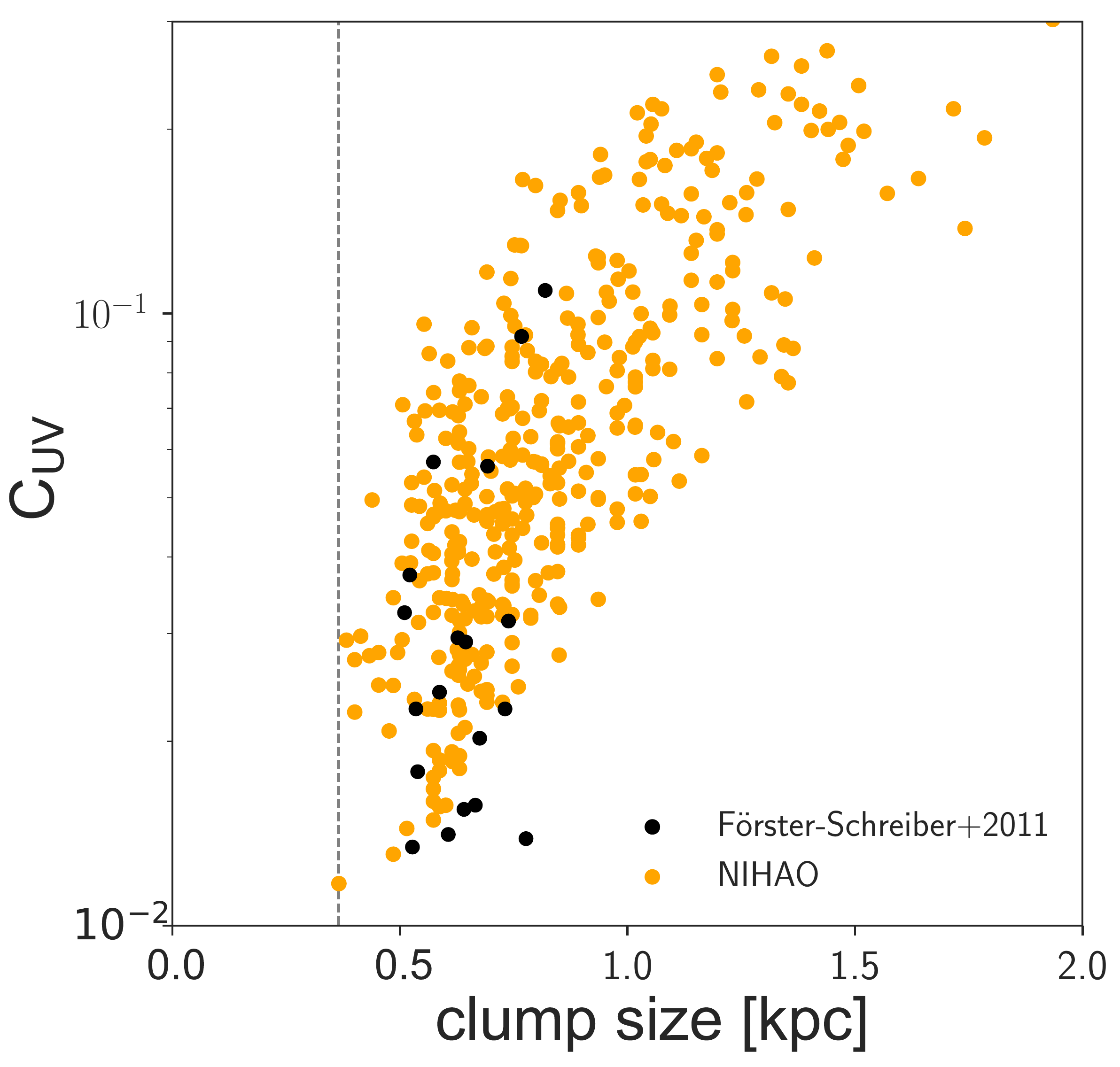}
\vspace{-.35cm}
\caption{Clump contribution to the rest-frame dust attenuated UV light of galaxies as a function of clump size for the 
NIHAO sample (orange dots) and observations from \protect\cite{Forster2011} (black dots). The gray dashed line indicates the lower limit on clump sizes set by our clump finding procedure.}

\label{fig:flux_frac_size}
\end{figure}

As first test we want to check how the clumpiness of the NIHAO galaxies
compares with observed galaxies, to assess if our galaxies have
a realistic number of structures in their light distribution.

In  order to  compare our  simulations with  the observations  used by
\cite{Guo2015},  we  constructed  HST  mock  images  with  {\texttt{GRASIL-3D}}
\citep{Dominguez2014}  for all  snapshots in the redshift range  $0.3<z<3$ of  our 19
NIHAO galaxies.
We then apply the same filters (F435W, F606W and F775W) and surface
brightness cuts used by \cite{Guo2015} to our mock images and look for clumps as described in section
\ref{ssec:RTclumps}.

In Fig. \ref{fig:clumpy_frac}  we compare  the fraction  of simulated
galaxies  with at  least one  off-center clump  with that of observations.
The  left panel shows  the comparison between  the complete
simulation sample (blue) and different observations.
Overall the clumpy fraction  in NIHAO galaxies 
follows quite well the clumpy fraction derived by \cite{Shibuya2016} (purple
dotted line) and agrees within the error bars with the results of 
of \cite{Guo2015} (black  short-long dashed line). 
The right  panel  shows  the results when galaxies are  separated into three stellar mass
bins     $\log(\rm      M_{*}/\Msun)<9.8$,     blue;
$9.8<\log(\rm M_{*}/\Msun)<10.6$,  green;  $\log(\rm M_{*}/\Msun)>10.6$,  red.

There is  a peak in the  clumpy  fraction  of  about  60\%-70\%  at  $z=1.5-2$  in  both
simulations and observations. At higher and lower redshifts the clumpy
fraction goes down  to 50\% at $z\sim3$ and to 40\%  at $z\sim0.5$,
dropping even further in the simulations to 20\% at $z\sim0.25$, thus
matching observations by \cite{Murata2014}.

When inspecting the different mass bins separately we still find a quite good agreement
between simulations and observations.
Specifically we see  that at  low redshifts  the two
less massive bins (blue and green bands and points) agree
very  well  with   observed  clumpy  fractions for the  same  mass
ranges. However, the  highest mass  galaxies (red  band) in  NIHAO show  a
clumpy  fraction slightly  too high  but still  in agreement  with the
observations within  their error  bars.
At intermediate  redshifts the high and
low  mass bins agree again well  with the observations
but  the  intermediate mass  bin (green band)  shows  a too  high  clumpy
fraction    although    in    agreement   with    measurements    from
\cite{Wuyts2012}.
Finally due to the selection function of the NIHAO galaxies,
we do  not have any data  for the highest mass bin, however,
the  other two  mass bins agree well with the observed clumpy fraction.

\begin{figure*}
\includegraphics[width=\textwidth]{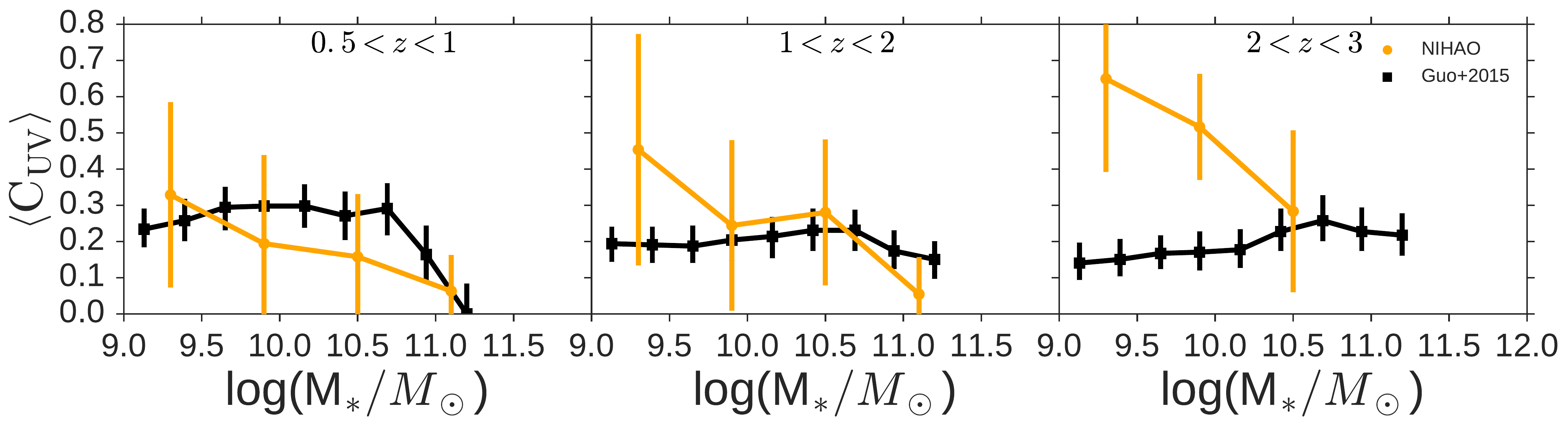}
\vspace{-.35cm}
\caption{Average intrinsic clump contribution to the rest-frame dust attenuated UV light of galaxies as a function of stellar mass in three different mass bins. The observations from \protect\cite{Guo2015} are shown as black squares and the results from NIHAO are shown with orange dots.}
\label{fig:conc}
\end{figure*}

Another observable  to which we can  compare  our RT  calculations  is  the
fractional contribution of the clump's UV flux to the total galaxy UV flux. Following \cite{Guo2015} we call this quantity
$C_{\rm  UV}$. In Figures  \ref{fig:flux_frac_size} and \ref{fig:conc} we show $C_{\rm  UV}$ as a function of the clump size and of the galaxy mass, respectively.

Clump sizes found in the RT calculations of the NIHAO sample are in agreement with observed clump 
sizes from \cite{Forster2011} and with intrinsic clump sizes if the pixel scale is matched to the HST pixel scale (see Appendix B).

Figure \ref{fig:conc} shows the comparison between the observed $C_{\rm UV}$ as a function of stellar mass
and the NIHAO one in three different redshift  bins  $0.5<z<1$,  $1<z<2$ and  $2<z<3$.
As in \cite{Guo2015} the  clump contribution to the  total UV flux of the galaxy is calculated for  all galaxies not
only the clumpy ones.
In the two lower redshift bins simulations (orange points) agree well with observations (black squares) within their error bars,
while the contribution of UV light from clumps of
intermediate mass galaxies in the highest redshift bin is too high in NIHAO.
A partial explanation of this excess could be related to the difference
in the sample size in this redshift bin: $\sim$70 simulation snapshots  vs
several    thousand   galaxies    in
CANDELS. Furthermore we  did not add noise to our  images which might
lead to  recovering more clumps  than the observers would  do. Despite
this  discrepancy NIHAO  recovers  quite well  the  observed UV  light
fraction  of clumps.

Mock observations of simulated clumpy galaxies have already been studied
before using different methods (see e.g. \cite{Genel2012,Moody2014,Tamburello2016}) and similar results
to ours were found. \cite{Genel2012} converted the SFR of one of their zoom-in cosmological galaxy (s224) to H$_{\alpha}$ flux, 
convolving it with a Gaussian of FWHM=0."17 and putting it at redshift $z=2.2$ to mimick SINFONI observations. 
They found that their short lived clumps are consistent with the SINFONI observations of galaxies at redshift $z=2.2$.
\cite{Moody2014} run the radiative transfer code \texttt{SUNRISE} \citep{Jonsson2006,Jonsson2010,Jonsson2010a} on their sample of 8 cosmological zoom-in simulated clumpy galaxies producing 
mock HST observations in four different filters. Unlike this work these authors additionally add noise to their images. They 
find qualitatively very similar results to the ones found here; a clump selection in longer wavelength maps results in lower clump counts 
compared to shorter wavelength maps. Furthermore these authors find that if
clumps are selected in mock observations, stellar mass maps or gas surface density maps the outcome shows vastly different clumps 
(see e.g. their figure 5). There is only a minority of clumps showing up simultaneously in two or more maps. 
This is also confirmed by \cite{Tamburello2016} who used the radiative transfer code \texttt{TRAPHIC}  \citep{Pawlik2008,Pawlik2011}
to create H$_{\alpha}$ maps of ionizing radiation of their non-cosmological simulations of galaxies. These authors add different levels of noise to their mock images and convolve the images with Gaussians with different values of FWHM to mimick different spatial resolutions. They find that the recovered properties of clumps strongly depend on the noise level and the spatial resolution. Clump sizes and masses can change by more than a factor of 2 depending on the sensitivity and the spatial resolution.

However, in this study we focus on the evolution of the clumpy fraction.
This is the first time that such an observational motivated comparison of the
clumpy fraction in the light distribution between observations and simulations has been
carried out. Our results show that the NIHAO galaxies have a realistic
light distribution suggesting that they are a good testbeds for a better understanding
of the origin and fate of these luminous clumps.
Therefore, in the next sections we analyze the physical properties of the
intrinsic clumps.

\subsection{Properties of the intrinsic clumps in NIHAO}
In order to better understand the physical properties
of the clumps, we analyze in detail the {\it intrinsic} clumps.

Selecting U-band luminous clumps results in a variety  of clumps with
different properties for  every galaxy and over a  vast redshift range
($0<z<3$).  In  total we find  682 u-band  clumps in 488 snapshots
in   the   selected    redshift   range (203 snapshots within $3>z>1$, 155 within
 $1>z>0.5$ and 130 within $0.5>z>0.25$).

\begin{figure*}
\includegraphics[width=.75\textwidth]{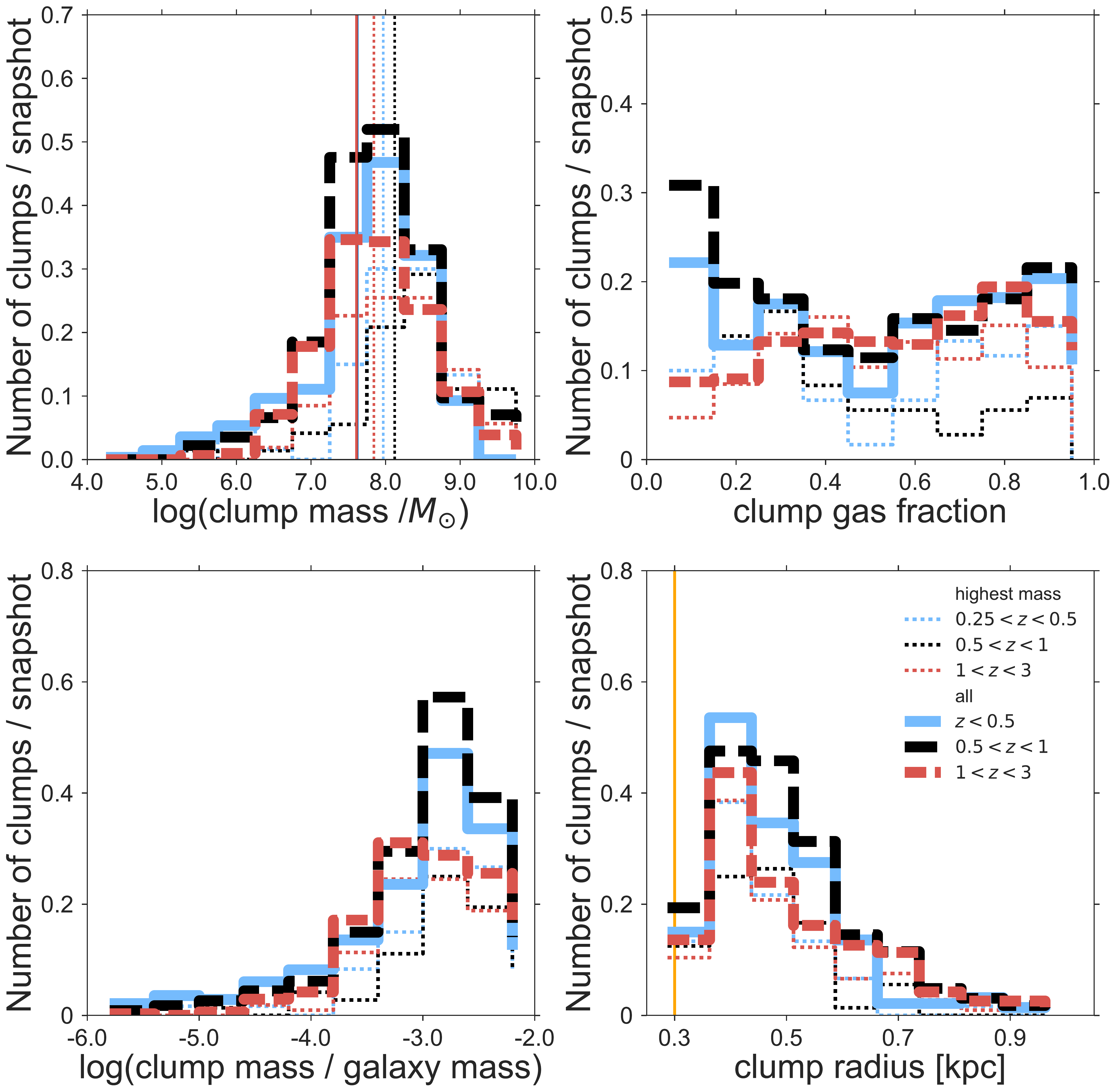}
\vspace{-.35cm}
\caption{Clump properties for three different redshift bins ($z<0.5$, $0.5<z<1$, $1<z<3$) for u-band selected clumps. \emph{The top left} panel shows the total clump mass, \emph{the top right}  shows the clump gas fraction, \emph{the bottom left} panel shows the clump mass as a fraction of galaxy mass and the \emph{bottom right} panel shows the clumps effective radii with the solid vertical orange line indicating the lower limit of clump sizes set by our selection criteria. Thick dashed lines show the results for the whole galaxy sample while thin dotted lines show results for only the 6 most massive galaxies. The colored solid lines in the top left panel show the median mass of clumps in the whole sample and the dotted lines show the median mass of clumps for the most massive galaxies.}
\label{fig:mass_func}
\end{figure*}

In Fig. \ref{fig:mass_func} we show some of the properties of NIHAO
UV \emph{intrinsic} clumps.
The mass (gas+stars) distributions is shown in the top left panel
of the figure.
Independently of  redshift the mass  of  clumps  in U-band  maps  always peaks  at
$\sim10^{8}\Msun$   with   a  maximum   clump mass   around
$\sim10^{9.5}\Msun$   and  a  minimum  around $\sim10^{6.5}\Msun$.
This minimum value is mainly due to the resolution of our simulation
which fixes the minimum pixel size of our maps (see section 
\ref{ssec:clumpselection}). These results do not change if we
 select only the 6 most massive galaxies which most closely resemble the observed galaxies as can be seen by comparing the thick dashed lines with the thin dotted lines in fig. \ref{fig:mass_func}.

The upper right panel shows the distribution of the clump gas fraction defined  as
$f_{\rm g}=M_{\rm  gas}/\left(M_{\rm gas}+M_{\rm  star}\right)$. Our clump 
finding algorithm has no specific lower mass limit. The lower limit on the clump mass is 
set by the size limit (300 pc) and the surface density of the selected clump. The lowest 
clump masses we find in the redshift bins are as follows: $0.25<z<0.5: 2.7\times10^4\Msun$, 
$0.5<z<1: 7\times10^3\Msun$, $1<z<3: 1.3\times10^5\Msun$
 The  clump gas fraction  shows a  slight  trend  to higher  gas  fractions at  higher
redshifts ($1<z<3$)  and a more  flat distribution at  lower redshifts
($0.5<z$). For intermediate redshifts we  see a slight bimodality with
clumps showing either  very low gas fractions or very  high ones close
to  $f_g=1$.
This  is in  contrast with  \cite{Oklopcic2016} who  find a
peaked distribution of clump gas  fraction with the maximum around gas
fractions  of  $f_g=0.3$. This discrepancy comes from the  difference  in  the
selection   method.    We   select   u-band   bright  clumps   while
\cite{Oklopcic2016} select clumps in gas surface density.

Despite the fact that  we find giant  clumps of masses greater  than $10^{8}\Msun$
the contribution  of these clumps  to the  total baryonic mass  of the
galaxy disc is  less than 1\% as shown in the lower left panel
of figure \ref{fig:mass_func}.
The  sizes of the clumps  found in the NIHAO galaxies
range from  $\sim$300 pc to  $\sim$900 pc with  a median
value of 450 pc for every redshift bin (see the Appendix for a comparison between observed and intrinsic clump sizes, as well as for the clump size dependence on pixel scale). The lower bound is  given  again
by our resolution (we fixed the minimum effective  radius  to be 300 pc).
Finally most of the clumps  we found are round(ish) in shape and sometimes
slightly elongated (see. e.g.  Fig. \ref{fig:surf}).

\begin{figure*}
\includegraphics[width=\textwidth]{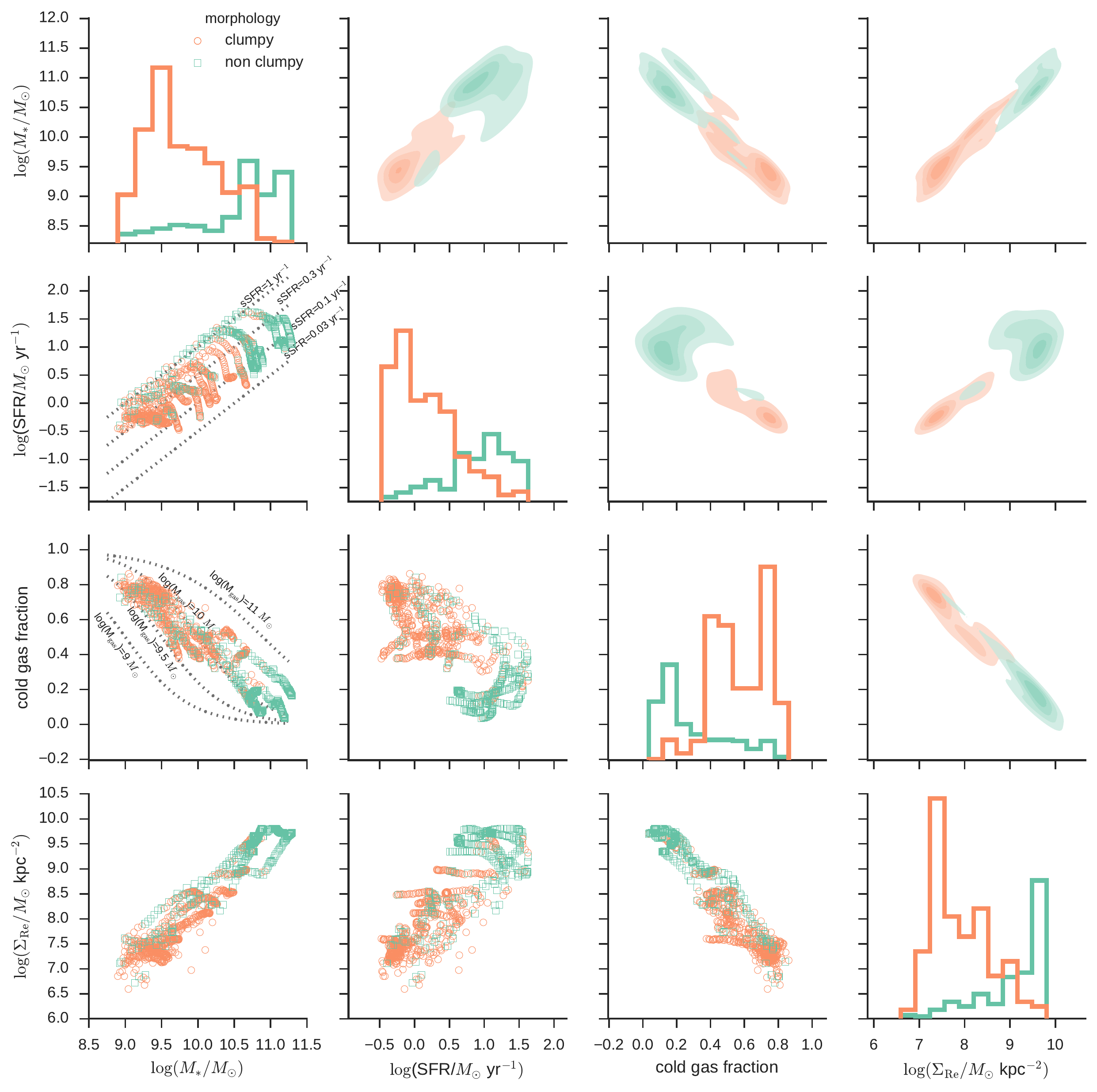}
\vspace{-.35cm}
\caption{Properties of host galaxies and their correlation with clumpy morphology for all galaxies and all redshifts in our sample. We show correlations of stellar mass (left column), SFR (second column), cold gas fraction (third column) and the mean surface density within the half mass radius (right column) with each of the other quantities and distinguish clumpy from non-clumpy galaxies. Red colored points and lines show galaxies with a clumpy morphology (at least one off center clump) and green squares/lines show non clumpy galaxies. The lower left triangle shows all snapshots for all galaxies with single dots while the upper right triangle shows a kernel-density estimation of the point distribution. The diagonal shows the marginal histogram of the property of the given column.}
\label{fig:corr}
\end{figure*}

The next question to address is what determines
if a galaxy has one or more luminous clumps.
In Fig. \ref{fig:corr} we  show  the correlation  between  stellar  mass,
SFR,  cold  gas fraction ($f_{\rm g}$) and the mean surface density  within the half
mass radius  ($\Sigma_{\rm Re}$) and color  code each galaxy
according to the ``morphology''  of the  light distribution (clumpy=orange,
non-clumpy=green).

In the  lower triangle  of the  plot every galaxy  is shown at every  snapshot
in  the redshift range $0<z<3$ as a point while in the upper right triangle we
show a  kernel density  estimation of the  point distribution.
On the diagonal  we show  the  marginal  histograms of  the  property in  the
according column for clumpy/non-clumpy galaxies.

Clumpy  and  non-clumpy  galaxies  mostly  separate  in  two  distinct
populations in these parameter spaces.  We find  that clumpy  galaxies show
high cold gas fractions, are  less centrally concentrated (lower value
of  $\Sigma_{\rm Re}$)  and  show  low and  average  SFRs and  stellar
masses.   In  contrast   non-clumpy   galaxies   are  more   centrally
concentrated, have  low gas fractions  and are among the  highest mass
galaxies with high SFR.
Qualitative  similar correlations are found by
\cite{Tadaki2014}  (e.g.   their  Fig.   1)  for  galaxies   from  the
SXDF-UDS-CANDELS field  and by \cite{Shibuya2016} (e.g.  their Fig. 7,
8,  9)  for  HST  photo-z  and Lyman  break  galaxies.

\begin{figure*}
\includegraphics[width=\textwidth]{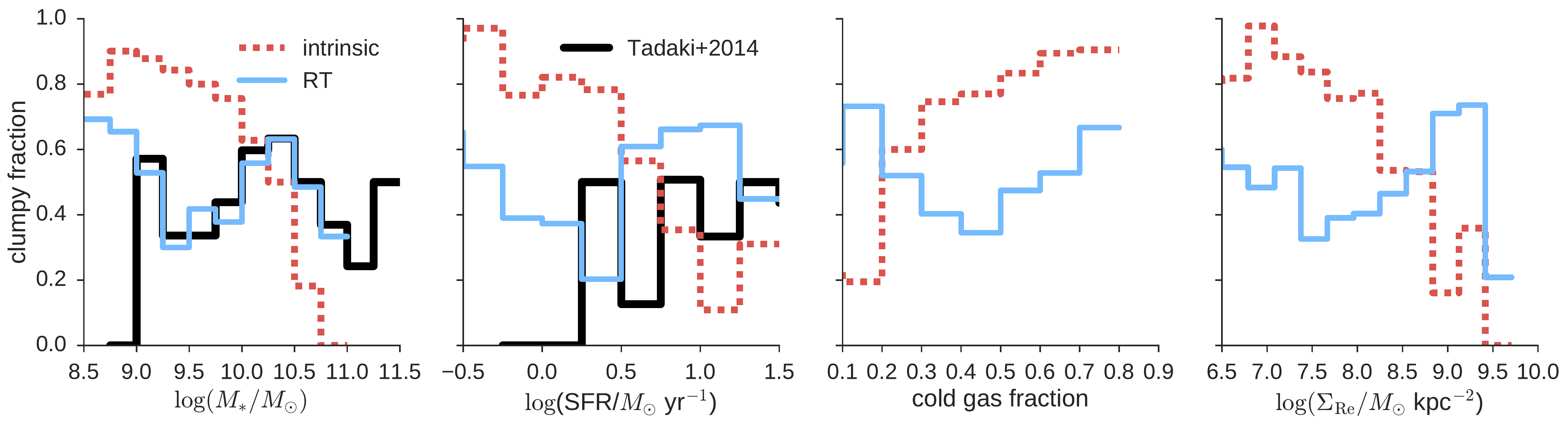}
\vspace{-.35cm}
\caption{The fraction of clumpy galaxies as a function of host galaxy parameters for \emph{intrinsic} and \emph{observed} clumps in comparison to observations from \protect\cite{Tadaki2014}. From left to right we show the clumpy fraction as a function of stellar mass $M_{\rm star}$, SFR, cold gas fraction and the mean surface density within the half mass radius.}
\label{fig:corr_comp}
\end{figure*}

There is one notable difference between simulations and observations.
From observations one would expect the highest mass, highest star forming galaxies to be clumpy. 
Here we find that the  lower mass, lower star forming galaxies are preferentially clumpy. 
This can be explained by the fact that our simulations are more centrally concentrated than observed galaxies, 
and as a consequence the gas in the simulated discs is less prone to gravitational instabilities \citep{Martig2009}.
Furthermore, we  use non  dust-attenuated properties for this plot. 
As we showed in Fig. \ref{fig:surf_GRASIL}, the degree of clumpiness in non-dust attenuated images can be different from that of dust-attenuated ones. 
Some of the most massive galaxies in particular, are not clumpy in the non-dust attenuated maps, but clumpy in the mock radiative transfer images.
However, other galaxies show the opposite behavior; they are clumpy in the non-dust attenuated maps and not clumpy in the mock images.
Therefore in figure \ref{fig:corr_comp} we give the clumpy fraction as a function
of stellar mass, SFR, cold gas fraction and mean surface density within the half mass radius for both \emph{intrinsic} and \emph{observed} clumps,
as well as the observational data from \cite{Tadaki2014}.
For our \emph{intrinsic} clumps we find an anti-correlation between clumpy fraction and stellar mass, 
SFR and mean central surface density and a correlation of clumpy fraction with the cold gas fraction of the galaxies.
While \emph{intrinsic} clumps show a strong evolution with all four galaxy parameters (anticorrelation with the galaxie's stellar mass, SFR and central surface density and correlation with the cold gas fraction) we do not find such an evolution 
for our \emph{observed} clumps. This is due to the beforehand mentioned dust attenuation.
Our \emph{observed} clumps, however, show very good agreement with the observational data, 
which also shows no strong correlation between the clumpy fraction and galaxy parameters.
This result further supports the need for a careful modeling of dust obscuration when comparing galaxy morphologies between simulations and observations.

\subsection{Clumps in light or clumps in mass?}

A lot of discussion is going on whether the observed clumps in high-redshift galaxies represent self-bound
clumps of stars orbiting within the disc or if they are simply a concentration of luminous young stars
with practically non dynamical influence on the disc.

\begin{figure}
\includegraphics[width=.95\columnwidth]{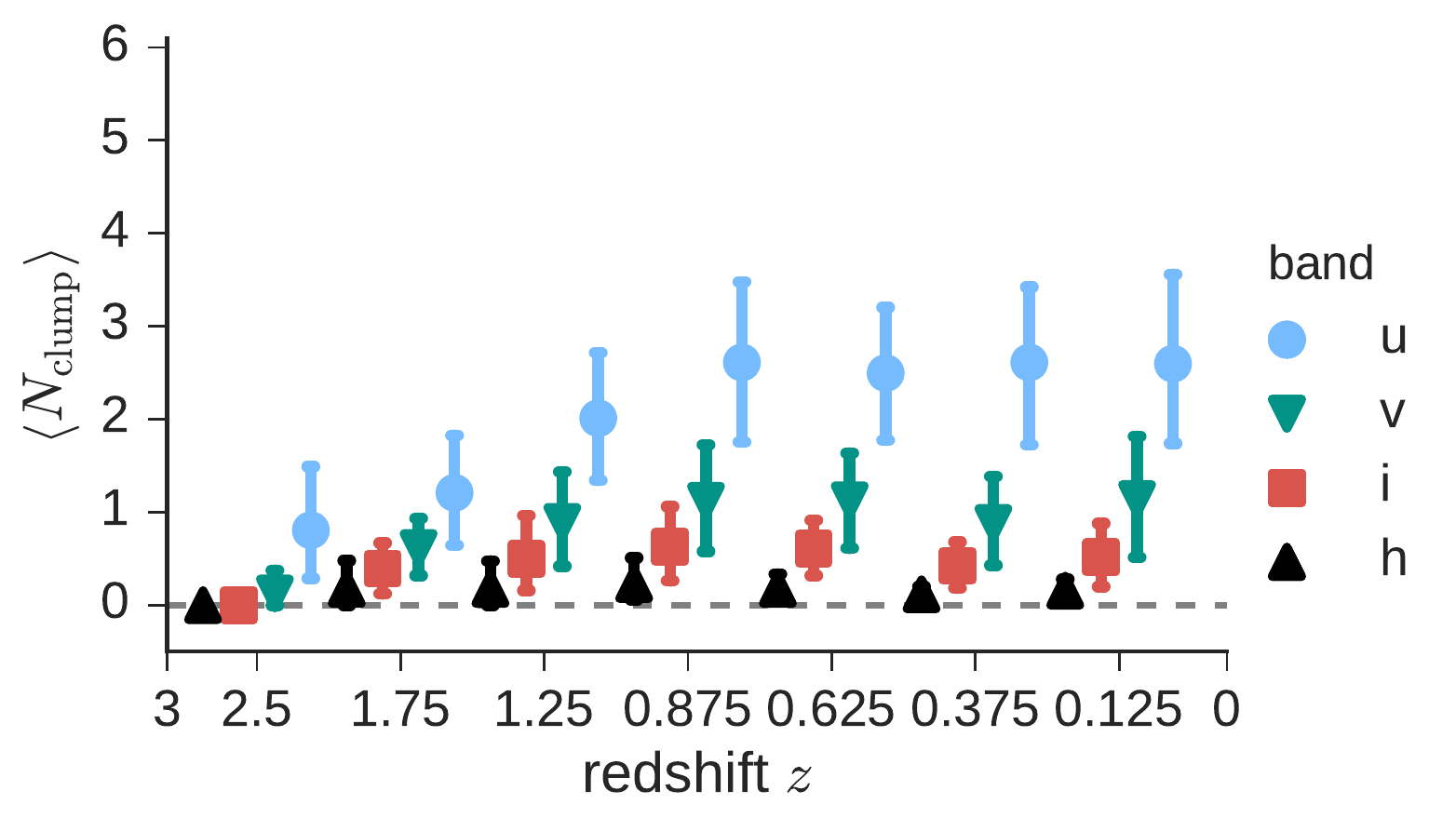}
\includegraphics[width=1.05\columnwidth]{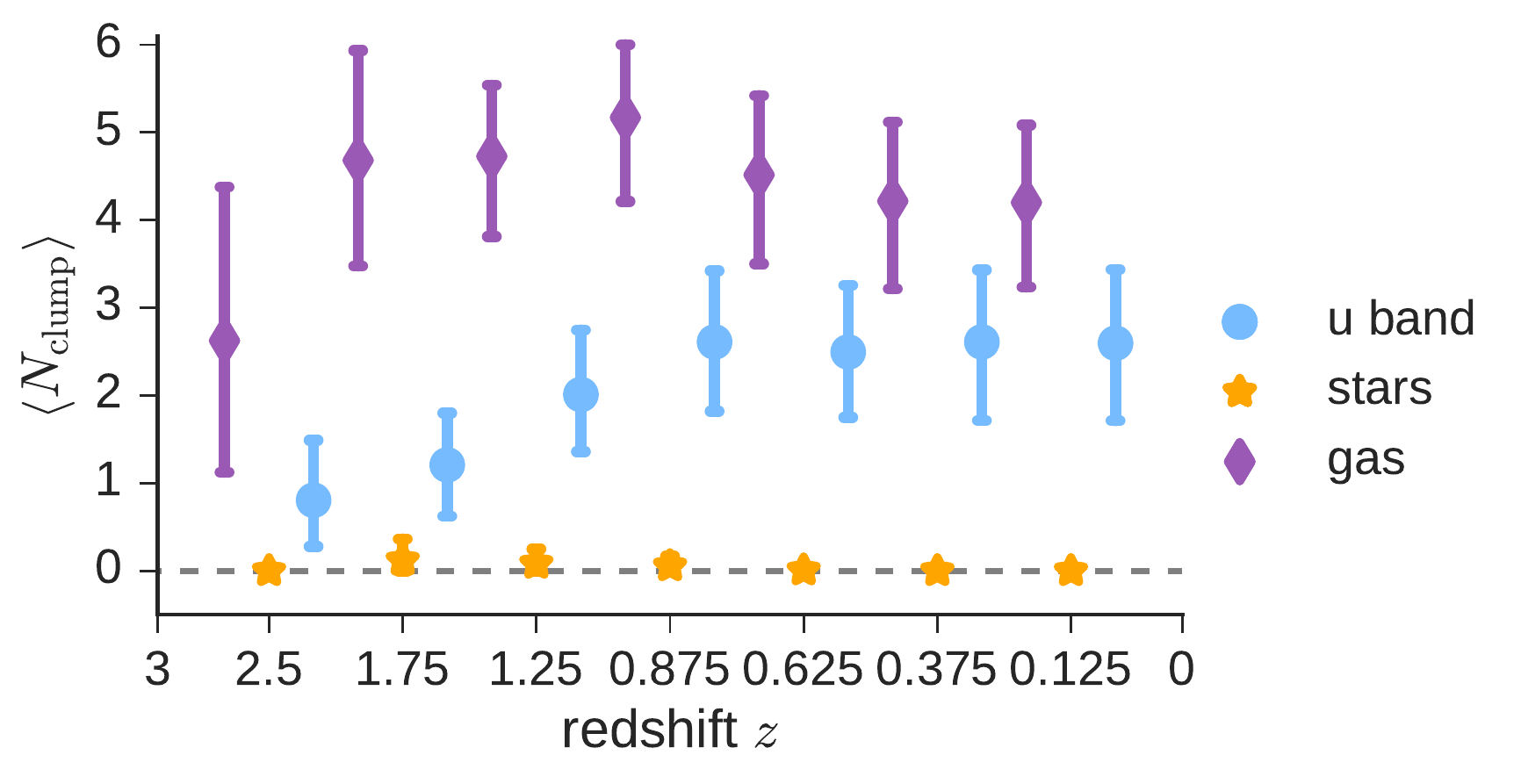}
\vspace{-.35cm}
\caption{Mean number of clumps per galaxy for clumps selected in different wavelength bands. The upper panel shows the mean number of clumps for clump selected in the u-band (blue dots), the v-band (green lower triangles), the i-band (red squares) and the h-band (black upper triangles). The lower panel shows a comparison of the mean number of clumps selected in the u-band (blue dots) with clumps selected in the gas surface density maps (purple diamonds) and the stellar surface density maps (orange stars). The points are slightly offset to avoid overlap.}
\label{fig:number}
\end{figure}

To start to answer this questions in the upper panel of Figure  \ref{fig:number}
we looked at the clumpiness of stars in different wavelength bands (u, v, i, h).
It is clear from the plot that moving to longer wavelength the disc clumpiness
tend to disappear, and practically no clumps are detected in the h band.

This seems to suggest that observed and simulated clumps in the non-dust attenuated u-band are
not actually bound structures of stars. This result is even more clear
in the lower panel of figure \ref{fig:number} where we look at
the presence (or lack thereof) of clumps in maps constructed
for different quantities: u-band luminosity, gas and stellar masses.
As already noted in figure  \ref{fig:surf} clumps in the u-band match
clumps in the gas surface density, but when we look at the stellar
mass maps they are extremely smooth and no clumps are detected
by our algorithm in practically any of our galaxies.

\begin{figure}
\includegraphics[width=0.5\textwidth]{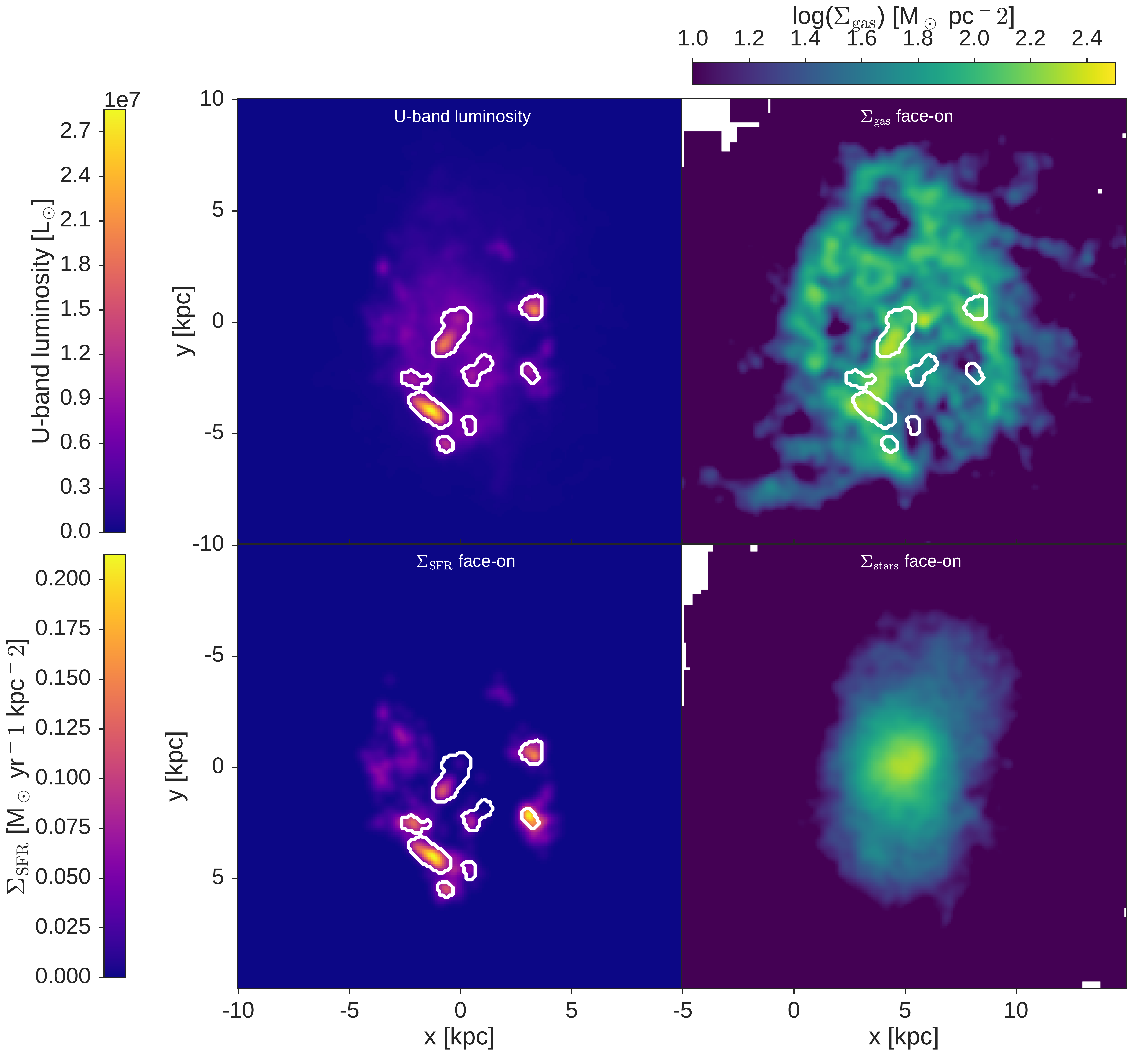}
\vspace{-.35cm}
\caption{Face-on maps  of u-band luminosity (upper  left), SFR surface
  density (lower left),  cold gas surface density  (upper right) and stellar mass surface density maps (lower  right). The stellar mass surface density map is extremely smooth. White contours show again the clumps selected in the u-band luminosity map.}
\label{fig:noclumps}
\end{figure}

This is confirmed by a visual inspection of one of the NIHAO galaxies
(same already shown in fig \ref{fig:surf}) in figure \ref{fig:noclumps},
where there is clearly a lack of any substructure in the stellar mass map.
This is maybe the most important result of our study. Despite
having galaxies as clumpy as the observed ones in {\it light} maps, we found
no evidence of any self-bound stellar structure in the {\it mass} maps
of the same galaxies.
We are facing here luminous clumps and not dynamical ones.
The good match between clumps in the u-band and the gas surface density maps
suggests that observed clumps are simply a manifestation of localized (clumpy) star formation regions,
as also observed in redshift zero galaxies.

\begin{figure}
\includegraphics[width=\columnwidth]{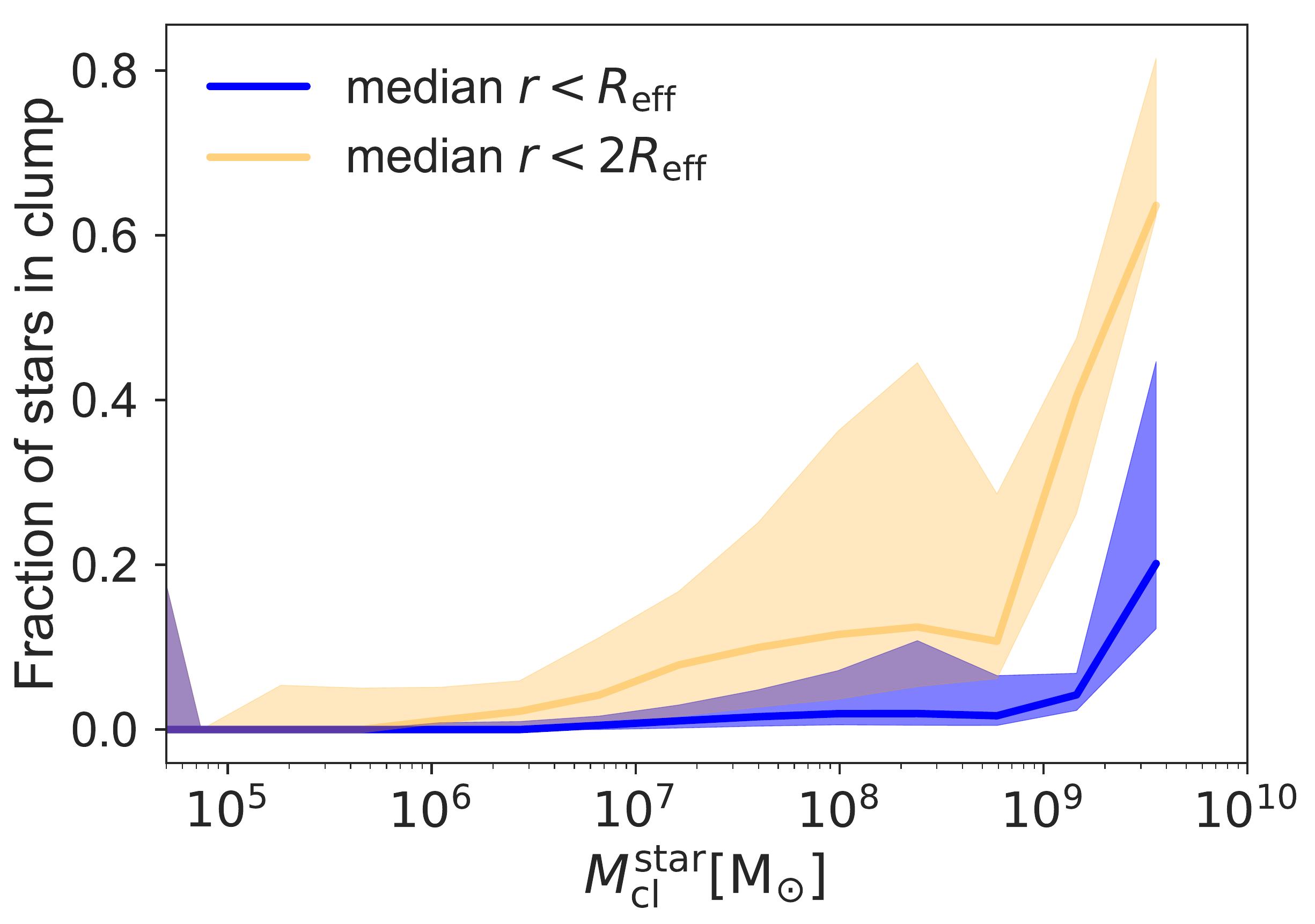}
\vspace{-.35cm}
\caption{Fraction of clump stars still close together after 200Myr. For every identified clump we show the fraction of clump stars still close together in the following snapshot. The blue line shows the median fraction of clump stars still within a region of 1$R_{\rm eff}$ around the clump's center of mass while the yellow line shows the median fraction of clumps stars still within 2$R_{\rm eff}$.}
\label{fig:frac_close}
\end{figure}

This is also confirmed by the life-time of the these luminous clumps: most of the clumps
disappear between two consecutive simulation snapshots in NIHAO, setting an upper limit
to their dissolution time of about 200 Myr, less than one dynamical
time of the galaxy. 
Figure \ref{fig:frac_close} shows the fraction of clump stars still close together in consecutive snapshots ($\sim$ 200 Myr). 
We track the star particles of every clump identified in a given snapshot via their particle IDs. In this manner, we evaluate how many stellar particles are still within a region of 1$R_{\rm eff}$ or 2$R_{\rm eff}$ around the clump’s center of mass in the next snapshots. We show the median fraction of clump stars still within a region of 1$R_{\rm eff}$ with a blue line and the according 16th and 84th percentile as a blue shaded region. The yellow line and yellow shaded region give the the median fraction of clump stars still within a region of 2$R_{\rm eff}$. With the exception of a few of the most massive clumps, we find that all clumps lose more than 90\% of their mass between two consecutive snapshots.

\begin{figure*}
\includegraphics[width=\textwidth]{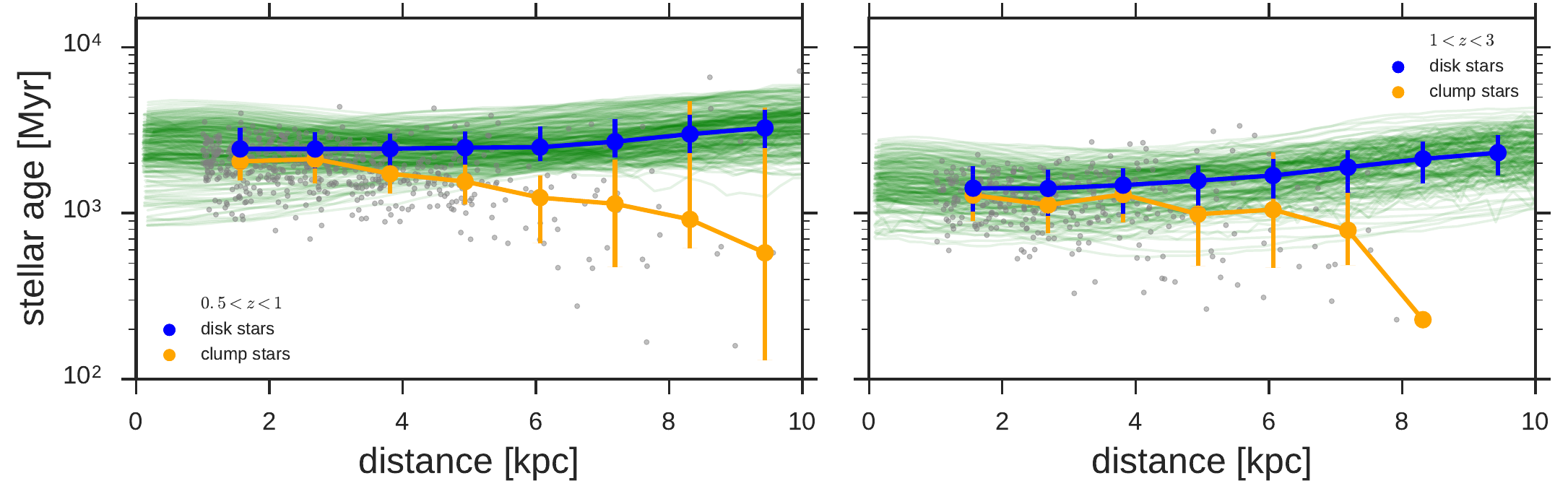}
\vspace{-.35cm}
\caption{Mean mass weighted stellar ages of individual clumps (grey dots) and the underlying disc stars (green lines) as a function of radius in two different redshift bins. The left panel shows the mean mass weighted stellar ages in the redshift range $0.5<z<1$ and the right panel shows the stellar ages for the redshift range $1<z<3$. The yellow dots show the median stellar age of individual clumps as a function of distance from the galaxy center and the blue dots show the median age of stars in the disc. The error bars show the 16th and 84th percentile.}
\label{fig:age}
\end{figure*}

\begin{figure*}
\includegraphics[width=\textwidth]{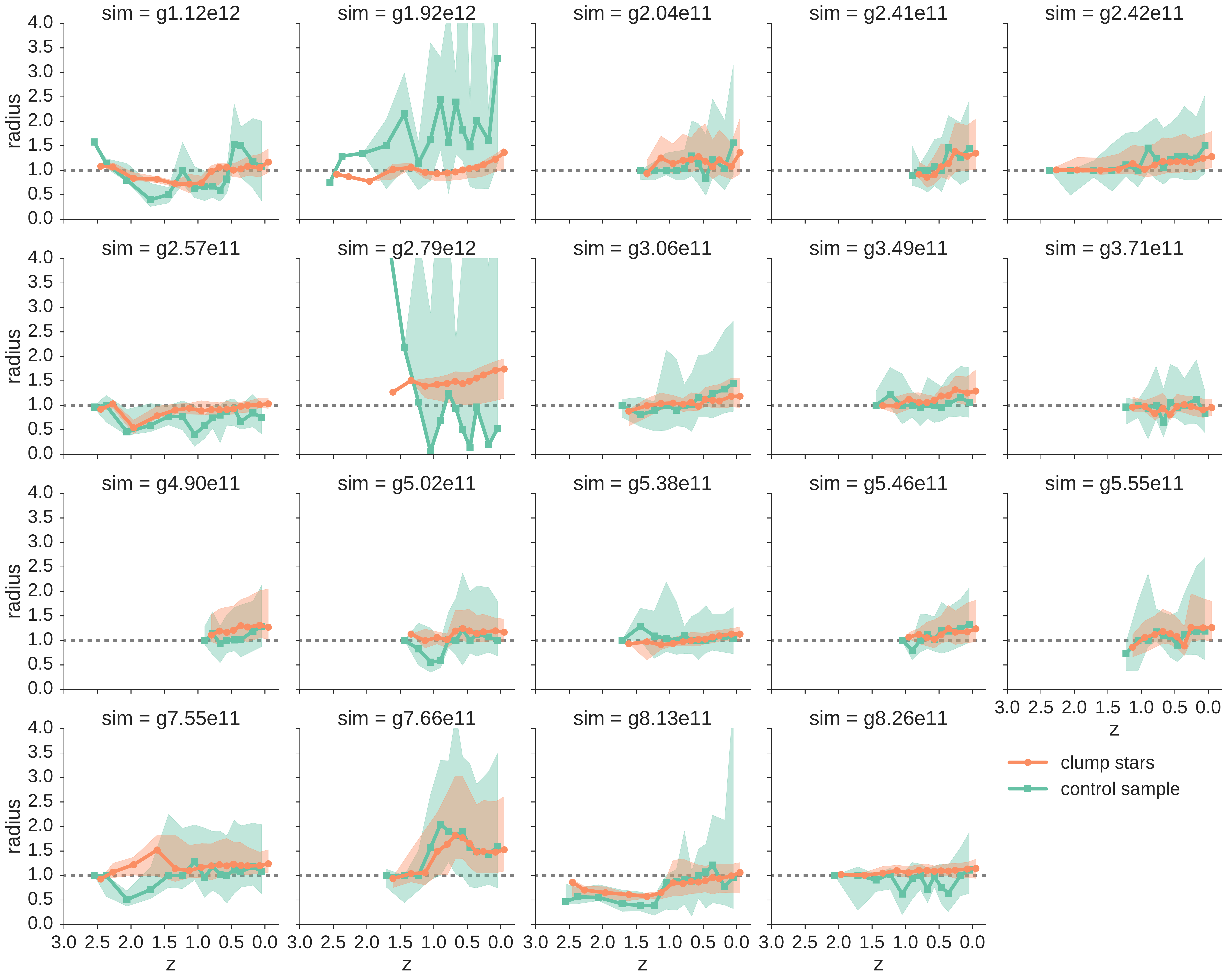}
\vspace{-.35cm}
\caption{Median distance of clump stars (red dots) and control stars (green squares) as a function of redshift. For every galaxy and for every clump we track the stars found to be in one clump at a given time and calculate their mean  galacto-centric distance at later times. We show the median value of all clump radii normalized to the initial clump radius. The same procedure is done for a control sample of stars of same number and same average radius as the clump stars. The shaded region shows the 16th and 84th percentile.}
\label{fig:inspiral}
\end{figure*}

These results are at odds with previous works \citep{Dekel2009a,Bournaud2009,Bournaud2014,Ceverino2010,Ceverino2015,Mandelker2014},
which suggested that clumps are gravitationally bound structures that
migrate towards the center under the influence of dynamical friction.
In support of this picture there is the observation of a color gradient of clumps as a function of distance from the galaxy center
 \citep{Forster2011, Shibuya2016}, with clumps closer to the center being redder
than outskirt clumps. The color gradient is supposed to be
a proxy for the clump age.
On the other hand it has already been suggested that such a gradient could simply be
due to the underlying stellar disc population \citep{Oklopcic2016,Nelson2016a,Genel2012}.

In figure \ref{fig:age}  we show the average age of simulated clumps as
a function of their distance from the center.
The stellar age of a clump is the (mass weighted) average age of {\it all} stars within the clump,
which of course is a mix of stars actually formed in the clump and the underlying stellar population.
As reference, we show in the same figure the mean stellar age of the global stellar disc
population as a  function of distance  from the  galactic center.
The gradient is quite weak with a slight increase  of stellar ages in the outskirts
of the disc, consistent with expectations from stellar migration \cite{Badry2016}.

We  find that  most clumps  are in the central parts of the galaxy within 5 kpc from the center,
(but keep in mind that we excluded  the innermost 1 kpc from our  analysis).
In NIHAO we find a wide range  of clump stellar ages from $\sim$200
Myr up  to about 2 Gyr  in the lowest redshift  snapshots in agreement
with  observed ages  of stellar  populations in  high redshift  clumps
\citep{Elmegreen2005, Elmegreen2009, Forster2011, Genzel2011, Guo2012,  Wuyts2012}.
Stellar ages of clumps  are generally slightly younger than the underlying mean  age of the disc  stars,
consistent with the picture of clumps being sites  of intense   star   formation

Interestingly we  recover the observed trend of
clumps in the outskirts being younger than clumps in the central parts
of the galaxies.
Contrary of what was suggested by some authors, who ascribe this gradient
to a migration of clumps, 
we find that this  effect is strongly due to a  selection bias.
Clumps  in the  central  parts of  the galaxy
include  more underlying disc  stars which increases  the mean stellar age  while 
in the  outskirts the  density of the stellar disc  decreases  and 
clumps  are  less polluted  by  disc  stars  and thus  appear younger.

Although we recover the observed trend of
clump ages as a  function of radius, this is not  a signature of clumps
spiraling inwards and moving to the  center as we will  show in the
next section.

\subsection{The evolution and final fate of light clumps}

If light clumps were self bound structures of stars
it would be natural to expect them to spiral in under the influence
of dynamical friction. Previous studies have indeed argued
that high redshift clumps can be responsible for the building of
bulges at high redshift \citep{Ceverino2010,Bournaud2014}.

As discussed in the previous section our simulations suggest a quite different scenario,
and hence it is interesting to ask what is the evolution of our luminous (but
not bound) clumps.

For every clump in every galaxy we track its evolution by calculating
the  mean  galacto-centric  distance of all its stars at  later  times ($r(z)$),
and we normalize this number by the distance at the time the clump was firstly detected ($r(z_{\rm form})$).
For stars moving inwards we expect the ratio $r(z)$/$r(z_{\rm form})$ to be below
one and the opposite for particles moving outwards.

Figure \ref{fig:inspiral} shows the evolution of the ratio
$r(z)$/$r(z_{\rm form})$ as a function of redshift for all our NIHAO galaxies.
In the plot we also show the same quantity for a ``control sample'' of disc stars;
for each clump with a given amount of stars, we select an identical  number of disc stars
at the same galacto-centric distance as the clump of interest at the time of its first detection.
The  shaded  region   shows  the  16th   and  84th
percentile. The  orange dots in Fig. \ref{fig:inspiral}
show that there is no preferential inwards migration for  clump stars.
If anything there is a  slight  trend for  clump stars  to  migrate outwards.
When compared to the control sample of disc stars, clump stars do not show any
particular difference, they behave in a very similar manner. 
Similar results for the lack of inward migration of clumps are found by \cite{Oklopcic2016} 
who looked at the angular momentum change between the final and initial snapshot in which a clump is visible.
Theses authors do find a roughly equal likelihood for clumps to lose and to gain angular momentum. 
Thus these results indicate that clump migration is not governed by dynamical friction but rather by gravitational torquing 
or tidal forces.

\begin{figure}
\includegraphics[width=\columnwidth]{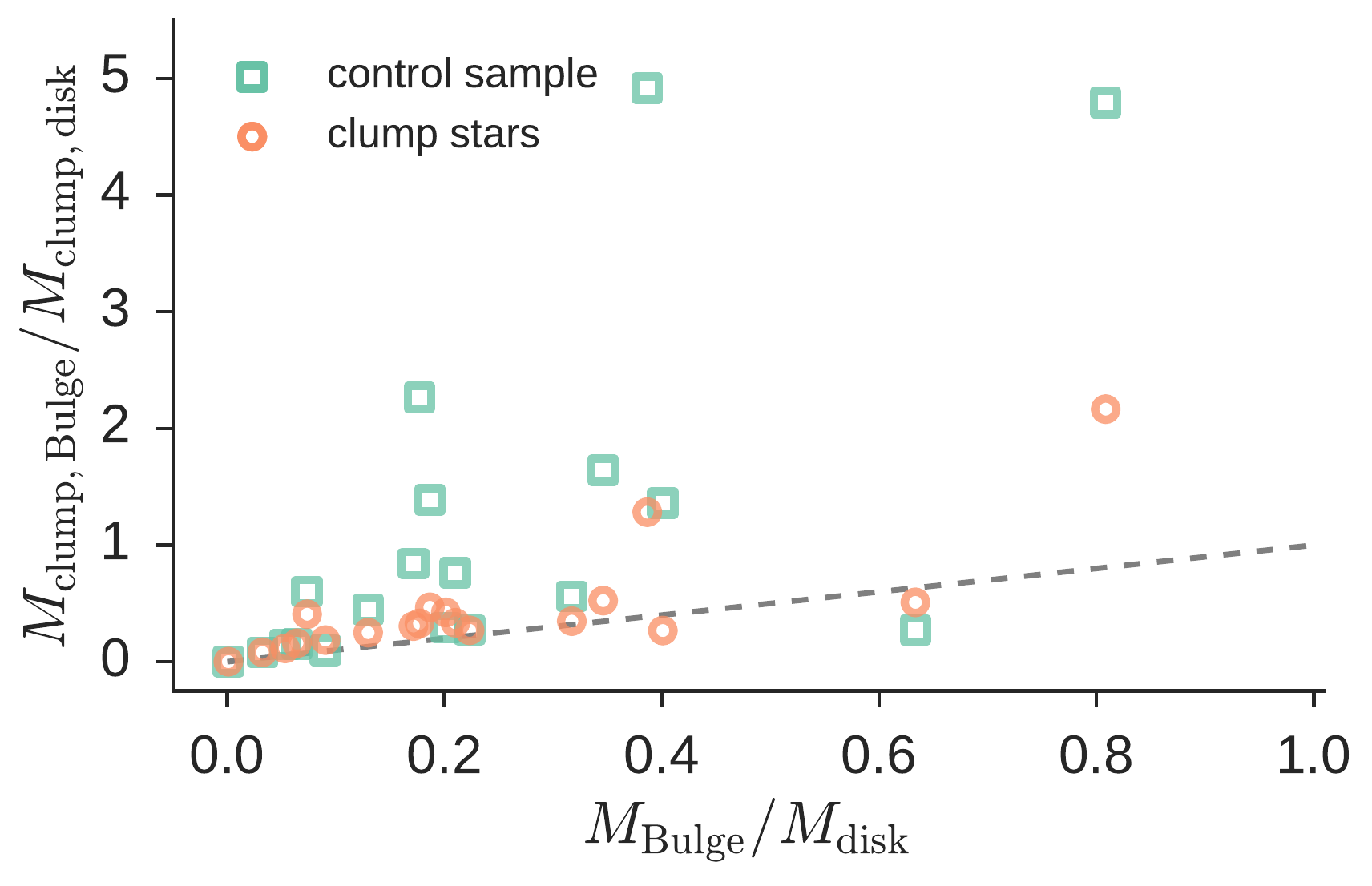}
\vspace{-.35cm}
\caption{Contribution of clump stars to the bulge and to the disc at redshift 0. Shown is the ratio of stellar mass from clumps in the bulge and in the disc as a function of bulge to disc ratio. The red dots show the result for stellar mass from clumps while the green squares show a control sample of disc stars chosen to have the same size and radius as the clumps stars.}
\label{fig:bulge}
\end{figure}

Since we plot the average distance, it could still be possible that a
substantial migration of stars inward  is compensated by a same amount of stars
moving outwards. Therefore, we  look in Fig.  \ref{fig:bulge} at
the mass contribution from clump stars to the galaxy bulge at $z=0$.
We first decompose  each galaxy  at redshift  zero into a disc and spheroidal component
using the same procedure described in \cite{Obreja2016}.
All stars marked as belonging to  the spheroid  and   having  a  galacto-centric distance
smaller than 1 kpc are regarded  as bulge stars.

We then calculate the fraction of disc and bulge mass due to stars that have been found 
in clumps at some earlier point in time.
In Figure \ref{fig:bulge} we plot the ratio of mass from clumps in the bulge $M_{\rm clump, Bulge}$
to mass from clumps in the disc $M_{\rm clump, disc}$ as a function of
bulge to disc ratio $M_{\rm  Bulge}/M_{\rm disc}$ of the galaxy.
Similar to the previous plot we show clump stars as orange dots and our control sample as
green squares.
The dashed grey line shows the 1:1 relation. Overall clump stars
seems to be equally distributed between  the bulge  and the disc at $z=0$.
Furthermore  when compared to the control  sample  clump stars
seems to have the same final fate as any other stars in the disc.

We can then conclude that, consistently with not being gravitational bound,
light clumps in high redshift galaxies do not preferentially
move inwards as time goes by and, do not preferentially contribute to  the bulge  growth.

\section{Summary}
\label{sec:summary}

We use 19 galaxies from the high mass end (M$_*>10^9$ \Msun at $z\sim1.5$) of the NIHAO sample to analyze them in detail for their clumpy morphology and quantify the clumpy fraction of this simulation suite in the redshift range $0<z<3$. The NIHAO sample is a suite of high-resolution cosmological hydrodynamical simulations of galaxies in the mass range $10^9\Msun < M_{200} < 4\times10^{12}\Msun$ which reproduces realistic galaxy properties over this huge range of galaxy masses. Unlike most other theoretical studies which looked for giant clumps in galaxies we do not select clumps in gas or stellar mass maps but we select clumps in luminosity maps. We apply two different selection methods: {\it intrinsic clumps} are selected in the non dust attenuated rest frame u-band images closely matching the observational selection method, while for {\it observed clumps} we run the radiative transfer code {\texttt{GRASIL-3D}} \citep{Dominguez2014}) on the 488 snapshots of our sample to create realistic mock HST observations of the galaxies. In this way we can directly compare the clumpy fraction of the NIHAO suite to the observed clumpy fraction from \cite{Guo2015} and \cite{Shibuya2016}.  

Our main findings can be summarized as follows:
\begin{itemize}
\item Comparing the observed clumpy fraction, the number of galaxies/snapshots with at least one off-center clump, of our RT images to the observed clumpy fraction of \cite{Guo2015,Shibuya2016} we find very good agreement between our simulations and the observations (see figure \ref{fig:clumpy_frac}). The NIHAO sample can well reproduce the observed number of clumpy galaxies and their evolution with redshift and perfectly matches the correlation of the clumpy fraction of galaxies with stellar mass (compare figure \ref{fig:corr_comp}). Sizes of clumps found in the NIHAO sample agree well with observed clump sizes (Fig. \ref{fig:flux_frac_size}) Furthermore, we recover the observed UV-light contribution of clumps to the total UV-light of the galaxy with $\sim$ 30\% of the light coming from clumps.

\item Selecting {\it intrinsic} clumps in the rest frame u-band images results in clump masses of few times $10^6\Msun$ to $10^9\Msun$, sizes of about 300 pc to 900 pc and gas fractions spanning a wide range from 0.1 to 0.9. These findings agree well with observed sizes and masses of giant clumps. However, although clumps are prominent in u-band luminosity maps they contribute only a small fraction of less than 1\% to the disc mass (compare figure \ref{fig:mass_func}).  

\item For this work {\it intrinsic} clumps are selected in stellar light and can only be found in young stars showing up in the the u-band. Selecting clumps in longer wavelength bands like the v-, h- or i-band the number of clumps found drops to zero (see figure \ref{fig:number}) and we can not find clumps in stellar mass at all. This recovers the findings of \cite{Wuyts2012} that clumps are only present in short wavelength images but not in the inferred stellar mass maps (compare also figure \ref{fig:noclumps}). Thus, we find clumpy star formation but no clumpy stellar discs in the NIHAO galaxies. 

\item Comparing the properties of \emph{intrinsic} clumpy and non-clumpy galaxies in figure \ref{fig:corr} we find a bimodality between these two. Clumpy galaxies show high cold gas fractions, are less centrally concentrated and show low and average SFRs and stellar masses. In contrast non-clumpy galaxies are more centrally concentrated, have low gas fractions and are among the highest mass galaxies with high SFR. These correlations for \emph{intrinsic} clumps are strongly altered if we use \emph{observed} clumps to divide our galaxy sample into clumpy and non-clumpy galaxies as was shown in figure \ref{fig:corr_comp}. Especially the clumpy fraction as a function of stellar mass for our \emph{observed} clumps is in very good agreement with the observed relation. Thus, we conclude that a careful modeling of dust obscuration has to be taken into account for a direct comparison of galaxy morphologies between simulations and observations. 

\item The mean mass weighted stellar ages of clumps in the simulations show the same trend as observed color gradients of clumps (e.g. \cite{Shibuya2016}). Clumps in the outskirts of the galaxies are younger and clumps in the center are almost as old as the underlying mean mass weighted stellar age of the disc stars (see figure \ref{fig:age}). This trend can be attributed to the fact, that the stellar density of disc stars is lower in the outskirts and higher in the center, thus when selecting clumps in images the contribution/pollution of disc stars to the clump is higher in the center than in the outskirts. That is why the mean mass weighted stellar ages of clumps and stellar disc are more similar in the galaxy center than in the outskirts.

\item We find that clumps in the NIHAO sample do not spiral inwards and do not contribute much mass to the bulge. Clumps get quickly disrupted and disperse. They lose about 90\% of their mass in less than 200 Myr (see fig. \ref{fig:frac_close}). We trace the stars of the clumps through time down to redshift zero and follow their mean radius and do not find any net inward migration of clump stars. Indeed, we find that clump stars and disc stars behave the same way, both do not show signs of a net inward migration as we have shown in figure \ref{fig:inspiral} and \ref{fig:bulge}. Furthermore, we quantify how much mass these stars contribute to the bulge and to the disc of the galaxies. Clump stars and randomly selected disc stars contribute the same mass to the bulge. Thus we do not see any indications that clumps in NIHAO would contribute to preferentially build up the bulge.
\end{itemize}

\section*{Acknowledgments}
We thank the referee for valuable comments that improve the readability of the paper.
The authors like to thank Hans-Walter Rix for fruitful discussions and very helpful comments on this work.
This research made use of the {\sc{pynbody}} package \citet{pynbody} to analyze the simulations and used 
astrodendro, a Python package to compute dendrograms of Astronomical data (http: //www.dendrograms.org/) to find clumps.
The authors acknowledge support from the
   Sonderforschungsbereich SFB 881 “The Milky Way System” (subproject
   A2) of the German Research Foundation (DFG).  Simulations have been
   performed on the THEO clusters of the Max-Planck-Institut fuer
   Astronomie at the Rechenzentrum in Garching and the HYDRA and DRACO cluster at the Rechenzentrum in Garching.
   Further computations used the High Performance Computing resources at New York University Abu Dhabi.
   We greatly appreciate the contributions of all these computing allocations.



\bibliography{astro-ph.bib}

\appendix
\renewcommand{\thefigure}{A\arabic{figure}}

\setcounter{figure}{0}
\vspace*{-.5cm}
\section*{Appendix:}

\subsection*{A: Sensitivity to detection threshold}
\begin{figure}
\includegraphics[width=\columnwidth]{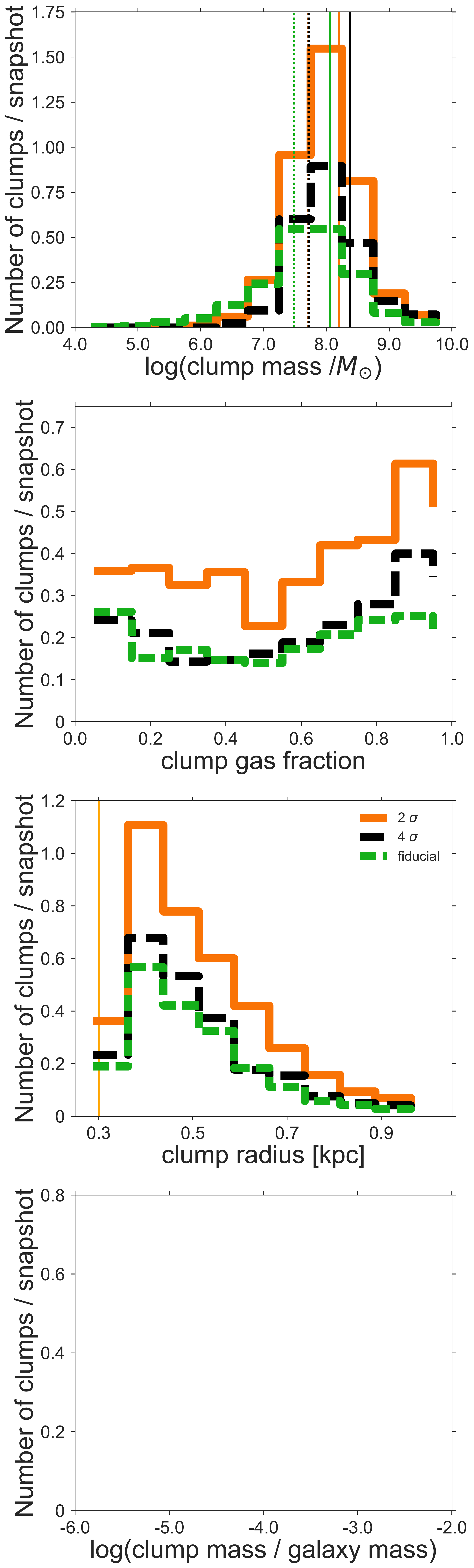}
\vspace{-.35cm}
\caption{Clump properties for three different clump selection thresholds (2$\sigma$, 3$\sigma$, 4$\sigma$) for u-band selected clumps. \emph{The top left} panel shows the total clump mass, \emph{the middle panel}  shows the clump gas fraction and \emph{the bottom left} panel shows the clumps effective radii with the solid vertical orange line indicating the lower limit of clump sizes set by our selection criteria. The colored solid lines in the top left panel show the median mass of clumps and the dotted lines show the mean mass of clumps.}
\label{fig:sensitivity}
\end{figure}

\begin{figure}
\includegraphics[width=\columnwidth]{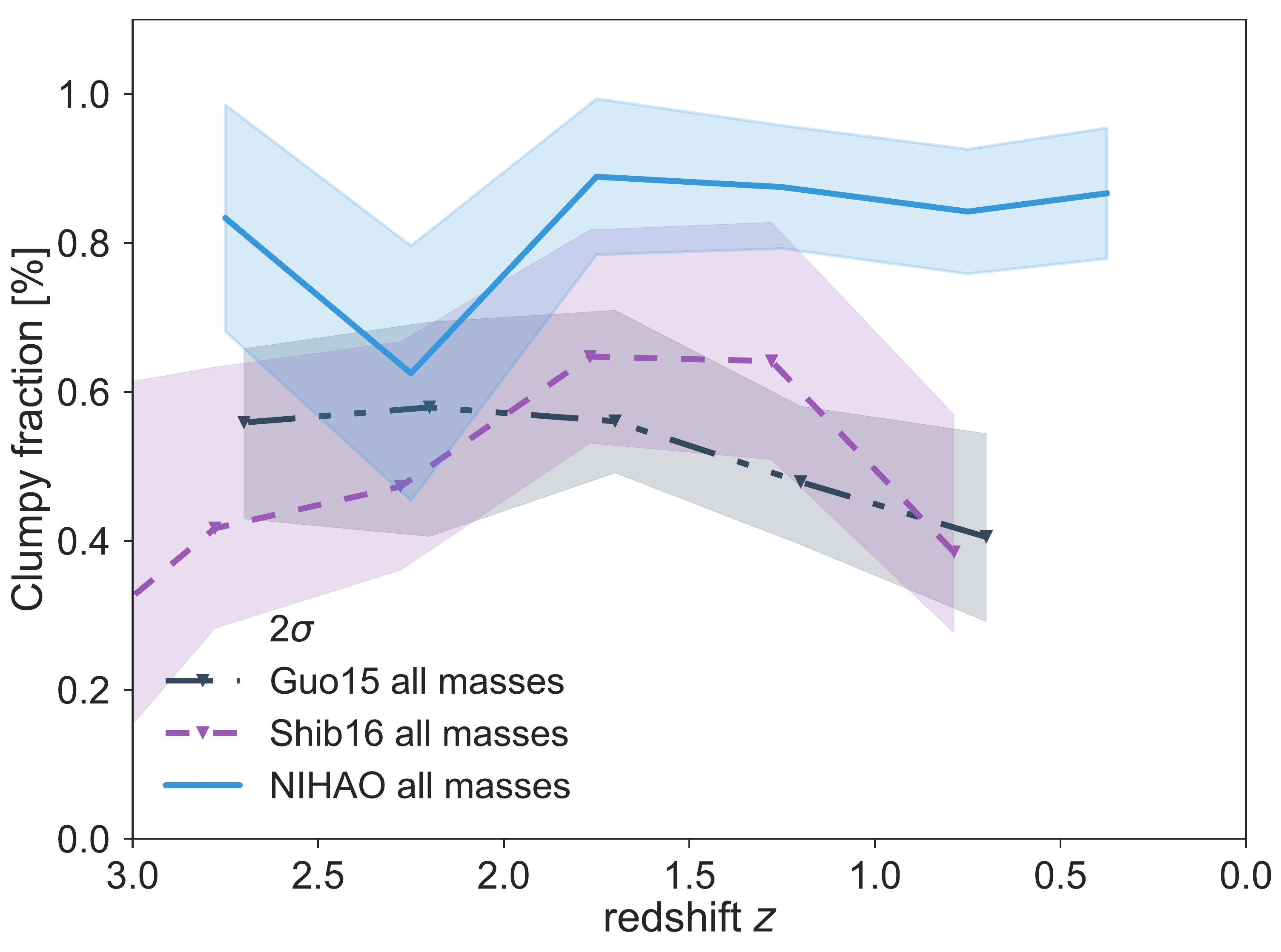}
\includegraphics[width=\columnwidth]{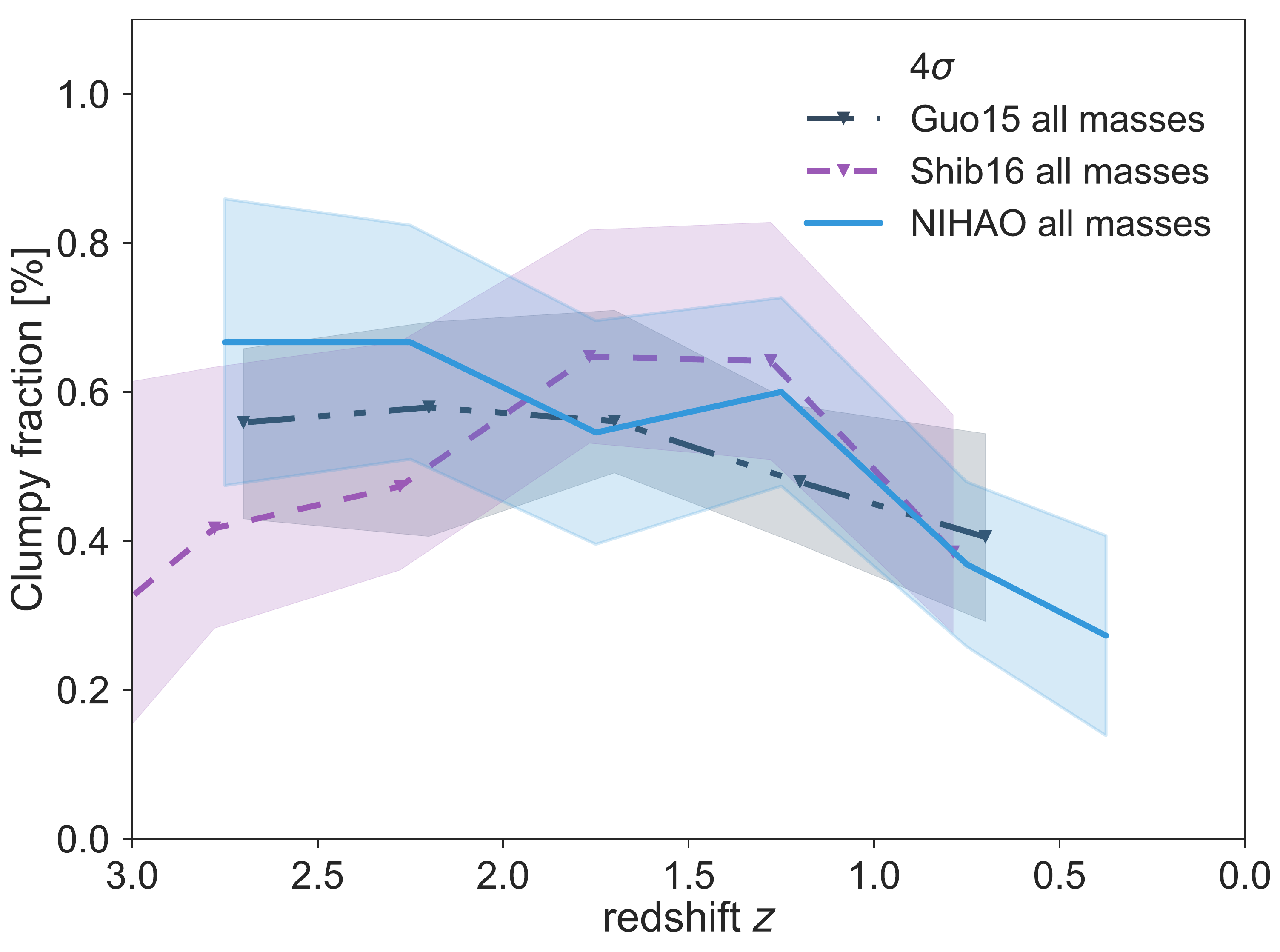}
\vspace{-.35cm}
\caption{Evolution of the fraction of galaxies with at least one \emph{observed} off-center clump for a selection threshold of 2$\sigma$ (upper panel) and 4$\sigma$ (lower panel). Both panel show the evolution of the clumpy fraction for the whole NIHAO sample (blue line) compared to observations from \protect\cite{Guo2015} (black dash-dotted line) and \protect\cite{Shibuya2016} (purple dotted line). The shaded band shows the $1\sigma$ scatter. }
\label{fig:rad_tr_comp}
\end{figure}

In this section we analyse how the properties of intrinsic clumps change if we change the sensitivity of our clump finding algorithm.
Our fiducial clump selection picks up clumps in the u-band images which stick out at least 3$\sigma$ above the mean u-band luminosity of the
10 kpc x 10 kpc images. In order to quantify the robustness of our results we rerun our clump finding algorithm on all our snapshots with a reduced threshold of 2$\sigma$ and with an increased threshold of 4$\sigma$.  In figure \ref{fig:sensitivity} we compare the clump properties found with all three thresholds. The upper panel shows the clump masses found, the middle panel shows clump gas fractions and the lower panel shows clump sizes. We do not see any differences in the properties of the clumps identified using different thresholds. Clump masses of all three selection thresholds peak at around $10^{8}\Msun$ However, we see as would have been expected that a clump selection with a threshold of 2$\sigma$ picks up much more clumps compared to the 3 and 4$\sigma$ selection. The same is true for the clump selection in the radiative transfer images. Figure \ref{fig:rad_tr_comp} shows the evolution of the clumpy fraction for a selection threshold of 2$\sigma$ above the mean flux in the given band (upper panel) and for a selection threshold of 4$\sigma$ (lower panel) as opposed to our fiducial selection threshold of 3$\sigma$ (compare left panel of fig. \ref{fig:clumpy_frac}). For a selection threshold of 2$\sigma$ above the mean flux we find a clumpy fraction which is higher than observed and for a selection threshold of 4$\sigma$ we find a clumpy fraction which is well in agreement with observations and the one obtained with a threshold of 3$\sigma$. By visual inspection of the clumps found we verify that a selection threshold of 2$\sigma$ picks up a significant amount of spurious, low surface brightness clumps in galaxies which by eye would not have been classified as clumpy. This is the reason for the increased clumpy fraction for this threshold value. The agreement between the threshold value of 3 and 4$\sigma$ makes us conclude that our fiducial selection threshold of 3$\sigma$ works properly.

\subsection*{B: Dependence of clump size on pixel scale}

\begin{figure}
\includegraphics[width=\columnwidth]{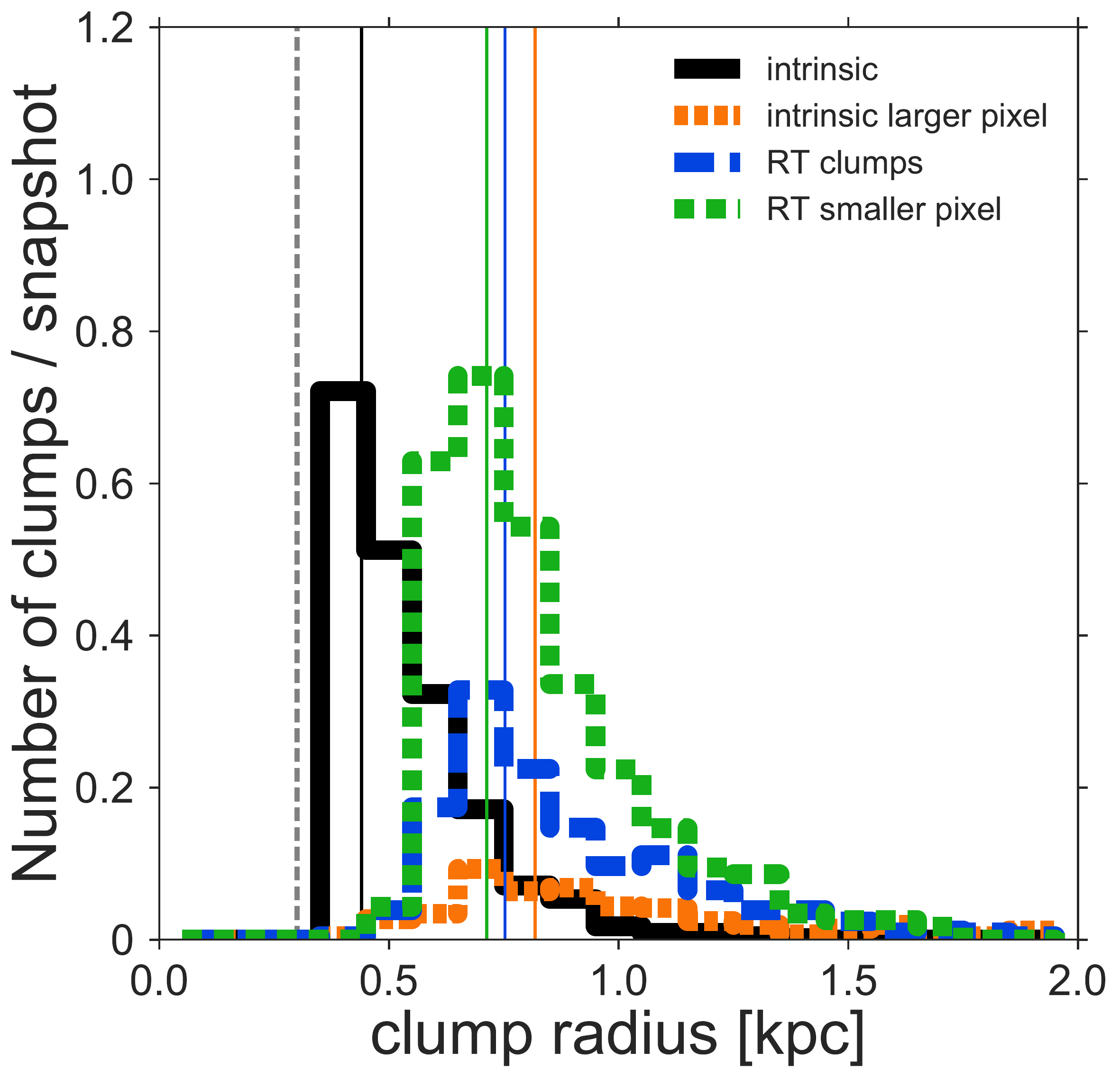}
\vspace{-.35cm}
\caption{Comparison of clump sizes for intrinsic clumps and clumps in the RT runs for different pixel scales. The black line shows our fiducial intrinsic clump selection, the orange line shows clump sizes for intrinsic clump selection with the HST pixel scale, the blue line shows clump sizes from our fiducial clump selection in the RT run and the green line shows clump sizes for RT runs with doubled resolution (half the pixel scale of HST). The colored vertical lines show the according median clump sizes and the vertical gray dashed line shows our resolution limit on clump sizes.}
\label{fig:size_comp}
\end{figure}

Recent work by \cite{Behrendt2016} and by \cite{Tamburello2016} found that clump sizes depend on the chosen spatial resolution and giant clumps would break up into smaller clumps if the spatial resolution is increased. In this section we test what happens to the clump sizes found in this study if the spatial resolution is changed. For our intrinsic clump selection we have chosen a pixel scale which is comparable to the physical resolution of our simulation. Therefore, we are already using the highest spatial resolution our simulations allow for. However, we change the pixel scale of our intrinsic clump selection to match the one of our RT runs (HST spatial resolution). Accordingly, we changed the minimum pixel per clump of our clump finder from 30 to 5 to keep the minimum effective radius of the clumps of about 300 pc. In Fig. \ref{fig:size_comp} we show the comparison of clumps sizes between the fiducial and the lower spatial resolution runs. We confirm that our clump finder picks up larger clump sizes in the lower spatial resolution runs, the median clump size increasing from  $\sim0.5$ kpc to  $\sim0.8$ kpc. From a visual inspection of the galaxies we find that this is due to two reasons: 1) some larger clumps in the lower spatial resolution run indeed break up into several clumps if the spatial resolution is increased and 2) most small clumps in the higher spatial resolution run are not found in the lower resolution run.

For the RT clumps we selected the 6 highest mass galaxies and rerun \texttt{GRASIL-3D} in the redshift range $z\sim3-1$ with a spatial resolution twice as large as that of the fiducial runs. Again we changed the minimum pixel per clump to match the physical resolution of the simulation. For these runs we adopted a value of at least 10 pixel per clump which results again in a minimum effective radius of $\sim300$ pc for the clumps. As figure \ref{fig:size_comp} shows there is almost no change in the clump sizes between the two different RT runs and clump sizes seem to be stable if the pixel scale is increased. It is unclear if doubling the spatial resolution is not enough to see a break up of clumps in the RT runs or if clumps do not break up in the RT run. We did not test the effect of increasing the resolution of the RT calculations because a resolution twice as good as the the fiducial one already requires four times more memory per core. In our case this would exceed the typical 2Gb available on HPC machines.
Finally, clump sizes of intrinsic clumps found in the images with HST pixel scale agree well with clump sizes found in the RT run of the same pixel scale.

\subsection*{C: Images}

\begin{figure*}
\centering
\includegraphics[width=\textwidth]{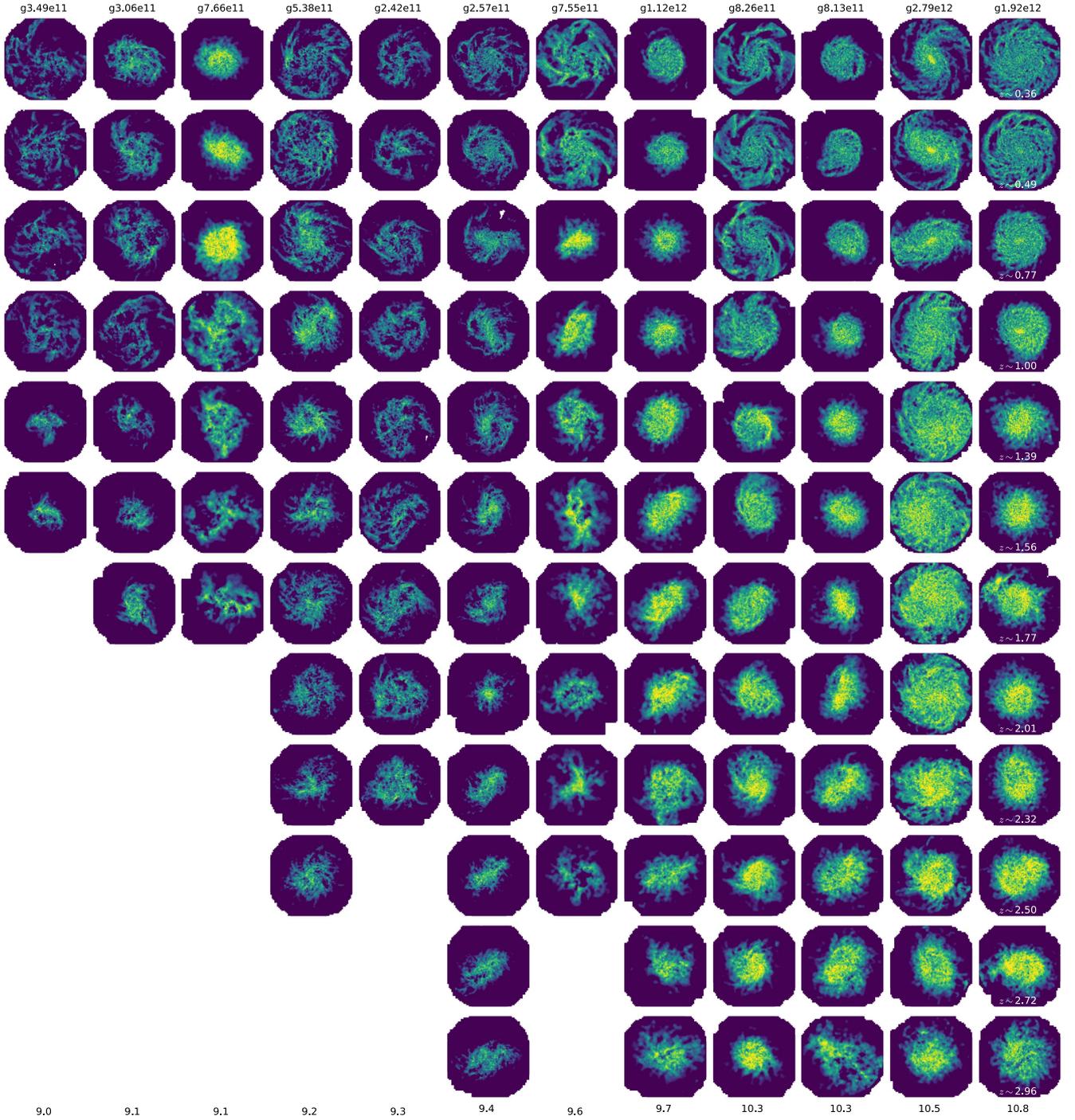}
\vspace*{-.5cm}
\caption{Impression of the cold gas surface density. From left to right: increasing stellar mass mass at redshift 0, from bottom to top increasing time.
Lower bounds of gas surface densities are set to 10\Msun pc$^{-2}$ and maximum values are set to 300\Msun pc$^{-2}$}
\label{fig:corr1}
\end{figure*}

\begin{figure*}
\centering
\includegraphics[width=\textwidth]{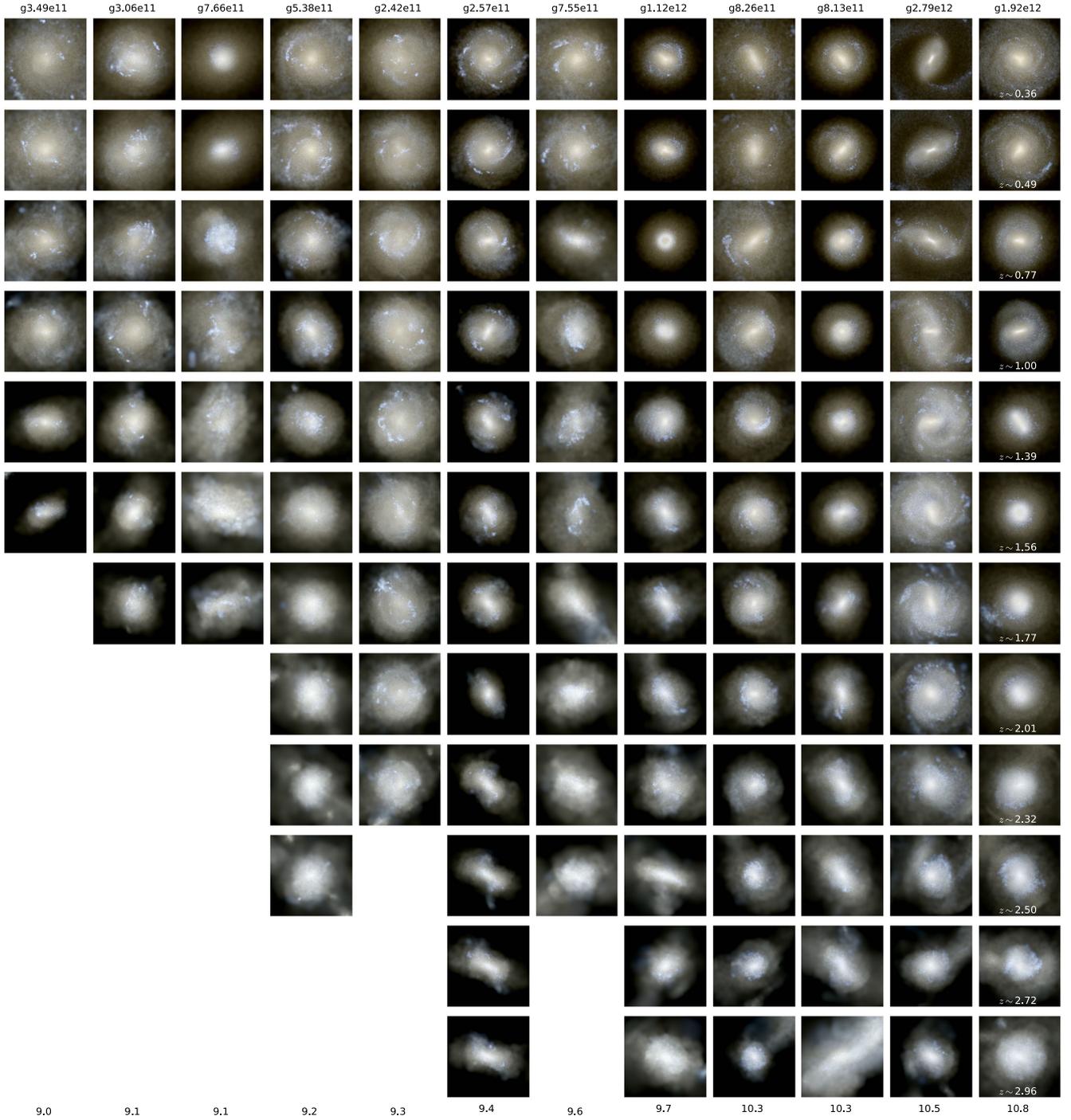}
\vspace*{-.5cm}
\caption{Impression of RGB composite images. From left to right: increasing stellar mass mass at redshift 0, from bottom to top increasing time.}
\label{fig:corr2}
\end{figure*}

\begin{figure*}
\centering
\includegraphics[width=\textwidth]{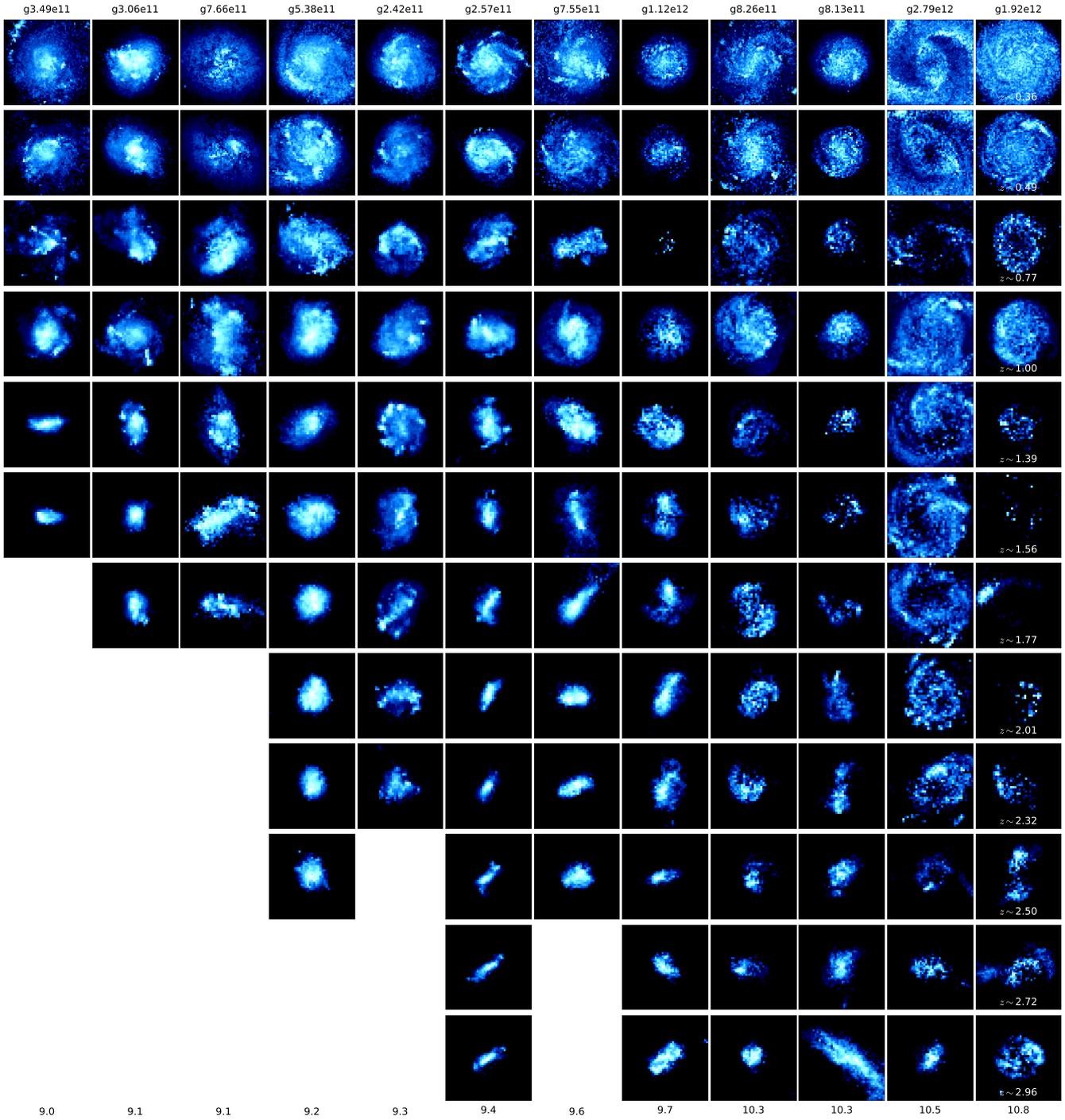}
\vspace*{-.5cm}
\caption{Impression of the outcome of the RT runs. From left to right: increasing stellar mass mass at redshift 0, from bottom to top increasing time.
For snapshots in  the redshift range  $3>z>2$ we select the  F775W filter,
for $2>z>1$  we select the  F606W filter and  for $z<1$ we  select the
F435W  filter   to  detect  clumps.  This   filter  selection  roughly
corresponds   to    the   rest    frame   UV   at    these   redshifts
\citep{Guo2015}.}
\label{fig:corr2}
\end{figure*}

\label{lastpage}

\end{document}